\documentclass[a4paper,11pt]{article}
\pdfoutput=1 

\usepackage{cancel}
\usepackage{jheppub} 
\usepackage[T1]{fontenc} 
\usepackage{overpic}

\usepackage{color}
\usepackage{amsthm}
\usepackage{graphicx}
\usepackage{slashed}
\usepackage{amssymb}
\usepackage{xspace}            

\def\slashchar#1{\setbox0=\hbox{$#1$}           
   \dimen0=\wd0                                 
   \setbox1=\hbox{/} \dimen1=\wd1               
   \ifdim\dimen0>\dimen1                        
      \rlap{\hbox to \dimen0{\hfil/\hfil}}#1 
   \else                                        
      \rlap{\hbox to \dimen1{\hfil$#1$\hfil}}/                                    \fi}

\title{\boldmath Jets from Jets:\\Re-clustering as a tool for large radius jet reconstruction and grooming at the LHC}

\author{Benjamin Nachman$^a$,}
\author{Pascal Nef$^a$,}
\author{Ariel Schwartzman$^a$,}
\author{Maximilian Swiatlowski$^a$,}
\author{and Chaowaroj Wanotayaroj$^b$}

\affiliation{$^a$SLAC National Accelerator Laboratory, Stanford University, 2575 Sand Hill Rd, Menlo Park,
  CA 94025, U.S.A.}
\affiliation{$^b$Center for High Energy Physics, University of Oregon, Eugene, OR 97403, U.S.A.}

\emailAdd{bnachman@cern.ch, pascal.nef@cern.ch, sch@slac.stanford.edu, swiatlow@slac.stanford.edu, ma.x@cern.ch}

\abstract{
Jets with a large radius $R\gtrsim 1$ and grooming algorithms are widely used to fully capture the decay products of boosted heavy particles at the Large Hadron Collider (LHC).  Unlike most discriminating variables used in such studies, the jet radius is usually not optimized for specific physics scenarios. This is because every jet configuration must be calibrated, insitu, to account for detector response and other experimental effects. One solution to enhance the availability of large-$R$ jet configurations used by the LHC experiments is {\it jet re-clustering}. Jet re-clustering introduces an intermediate scale $r<R$ at which jets are calibrated and used as the inputs to reconstruct large radius jets. In this paper we systematically study and propose new jet re-clustering configurations and show that re-clustered large radius jets have essentially the same jet mass performance as large radius groomed jets. Jet re-clustering has the benefit that no additional large-R calibration is necessary, allowing the re-clustered large radius parameter to be optimized in the context of specific precision measurements or searches for new physics.

}

\begin{document} 
\maketitle
\flushbottom

\section{Introduction} 

The angular separation of decay products for massive particles $\mathcal{P}$, such as $W$ and $Z$ bosons, scales as $2 m_\mathcal{P}/p_T^\mathcal{P}$.  This suggests that the radius parameter $R$ of jet clustering algorithms aimed at targeting the hadronic decays of $\mathcal{P}$ should be process dependent and scale with the momentum under consideration.  However, it is traditionally the case at the Large Hadron Collider (LHC) that only a few choices of $R$ are used for all analyses, which use the best available algorithm (which may not be optimal). This is because every jet configuration, which includes the algorithm, radius, and grooming parameters, must be calibrated to account for unmeasured energy deposits and other experimental effects~\cite{jescms,jesatlas}, even though the inputs to jet clustering (topological clusters for ATLAS~\cite{topo1,topo2} and particle flow objects for CMS~\cite{pflow1,pflow2}) are themselves calibrated. The calibration of inputs provides a partial calibration to the jet, but jet energy and mass scale corrections provide a \textit{full} calibration by also correcting particles that were missed, merged, or below noise thresholds, energy loss in un-instrumented regions of the calorimeter, and additionally takes into account correlations between particles. The dependence on these additional calibrations thus makes it desirable to reconsider the current jet clustering paradigm in favor of a modular structure that allows for a much broader class of algorithms and radius parameters to be selected by analyses.


One solution is to introduce a new angular scale $r<R$, such that jets of radius $r$ can be the inputs to the clustering algorithm of large radius $R$ jets\footnote{Similar ideas have been proposed in the past such as variable $R$ jets~\cite{variableR}.  While these methods address the variability of $R$, they do not address the concerns of individually calibrating many jet collections.}.  If chosen appropriately, the fully calibrated small radius jets can make the calibration of the re-clustered large radius jets automatic.  Furthermore, with no additional calibration needed, any large radius $R$, any clustering algorithm, and many grooming strategies can be used.  Using optimal parameters can, for instance, significantly improve the discovery potential of searches for new physics~\cite{jetography}.  In particular, every kinematic region of every analysis for every data-taking condition (e.g. level of pileup - number of additional $pp$ collisions) can optimize these parameters in order to maximize the sensitivity to particular physics scenarios.  Another benefit is that the uncertainties on the re-clustered $p_T$ and mass are also automatic consequences of propagating the corresponding uncertainties computed for small radius jets.  In this way, the re-clustered jet mass can be viewed as any other kinematic variable, such as di-, tri-, or multi-jet invariant masses that are ubiquitous in measurements and searches for new physics. The uncertainties on re-clustered jets may be higher or lower than those on the corresponding directly constructed large radius jet: a full experimental assessment will be necessary to determine if any precision is lost.

The idea of re-clustering small radius jets is not new.  These objects first appeared in an ATLAS search for supersymmetry in the multijet final state~\cite{atlasre-clustered} and more recently in an ATLAS search for direct stop quark pair production in the all hadronic final state~\cite{zerolepton}.  There are also related techniques which group small radius jets together to form pseudo-jets~\cite{cmspseudo} or mega-jets~\cite{cmsrazor}.  However, these analyses still use only a small number of re-grouping techniques which can be further optimized to depend on event kinematics.  The purpose of this paper is to describe a systematic comparison of re-clustering jet techniques.  Section~\ref{sec:description} describes the technical details of re-clustering, including re-clustered grooming which has never been considered before.  Section~\ref{sec:simulation} gives the simulation details for the comparison of  large radius jets and re-clustered jets presented in Section~\ref{sec:analysis}.  Finally, Section~\ref{sec:conc} provides some concluding remarks.

\section{Jet re-clustering}\label{sec:description}

The  clustering algorithms at the LHC use sequential recombination.  Given distance metrics $d_{ij}$ (between particle 4-vectors) and $d_{iB}$ (between particle and beam 4-vectors), these algorithms recursively combine proto-jets until there are none left.  The list of proto-jets is initialized by the set of jet inputs and then at every level of recursion, the algorithms combine particles based on $d_{ij}=\text{argmin}_{i',j'}\{d_{i'j'},d_{i'B}\}$.  If $j\neq B$ then proto-jets $i$ and $j$ are combined into a new proto-jet with a four-vector that is the sum of the four vectors of $i$ and $j$\footnote{This is a four-vector recombination procedure; other schemes also exist.}.  If $i=B$, then the proto-jet $i$ is declared a jet and removed from the list of proto-jets.

The most widely used metrics come from the $k_t$ family: $d_{ij}=\min(p_{Ti}^{2n},p_{Tj}^{2n})R_{ij}^2/R^2$, where $R_{ij}=\sqrt{\Delta y_{ij}^2+\Delta\phi_{ij}^2}$. For $n=1$ the algorithm is called $k_t$~\cite{kt}, for $n=0$ its is called Cambridge-Aachen (C/A)~\cite{CA}, and for $n=-1$ the algorithm is known as anti-$k_t$~\cite{antikt}.  In all three of these algorithms, $R$ is roughly the size of the jet in $(y,\phi)$ space, though C/A and $k_t$ jets can have very irregular {\it jet areas}~\cite{areas}.

\begin{figure}[htbp!]
\begin{center}
\includegraphics[width=.95\textwidth]{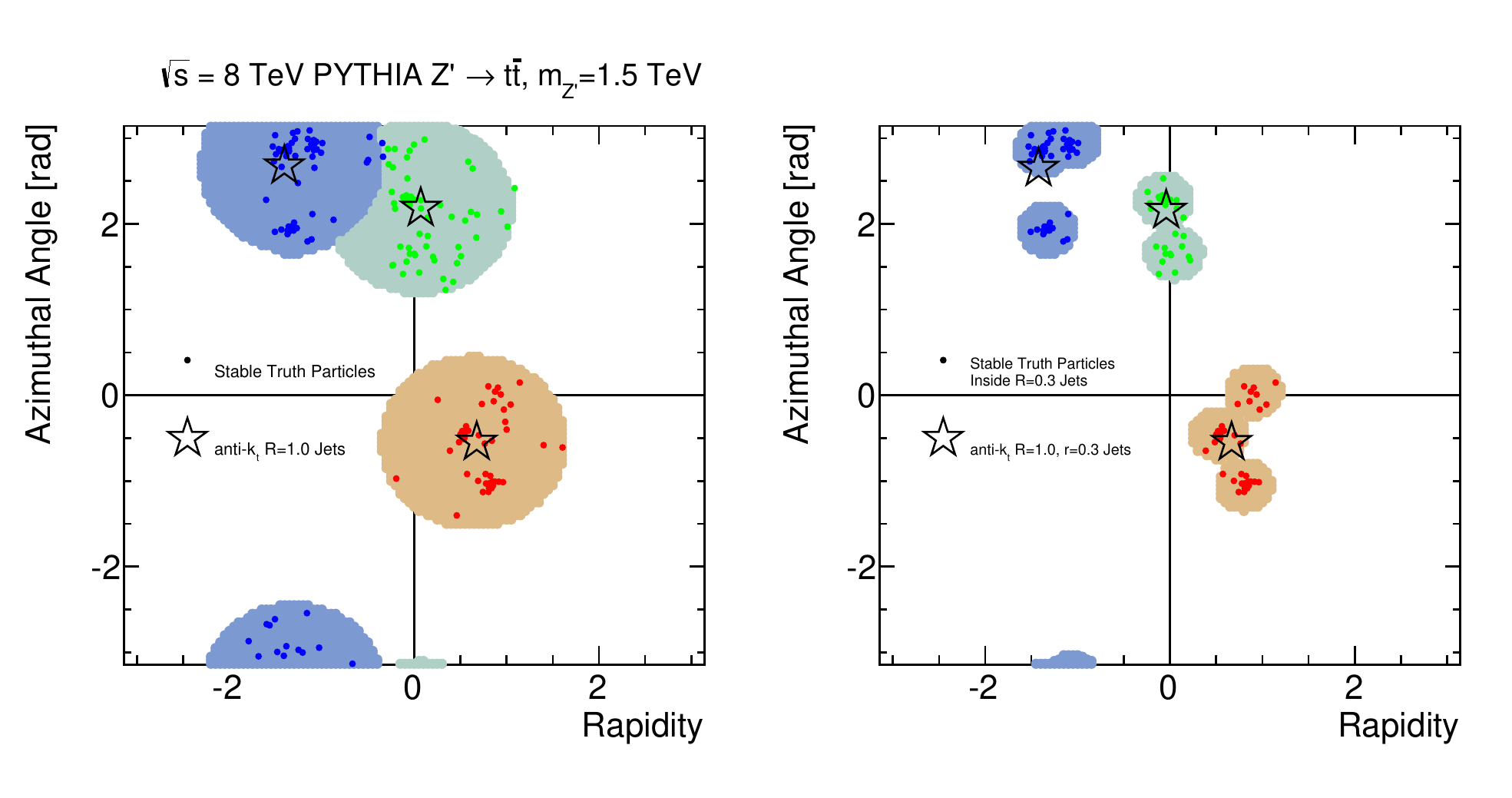}
\end{center}
\caption{An example event which has been clustered using the  anti-$k_t$ $R=1.0$ (left) and with anti-$k_t$ $R=1.0$ re-clustered $r=0.3$ anti-$k_t$ jets (right).  The shaded regions show the jet area determined by clustering ghost particles.  Only large radius jets with $p_T>50$ GeV are shown and small radius jets are required to have $p_T>15$ GeV. }
\label{fig:eventdisplay}
\end{figure}

The inputs of jet a clustering algorithm are typically stable particles (Monte Carlo truth studies), topological clusters (ATLAS), or particle flow objects (CMS).  Re-clustered large radius $R$ jets take as input the output of the  small radius $r$ jet clustering.  In general, the algorithm used to cluster the small radius jets can be different than the algorithm used for re-clustering the entire event.  Fig.~\ref{fig:eventdisplay} shows a simple example of an event clustered with  anti-$k_t$ $R=1.0$ and with anti-$k_t$ $R=1.0$ re-clustered $r=0.3$ anti-$k_t$ jets.  Unlike the inputs of  clustering which are e.g. measured in a calorimeter and can be measured and individually calibrated with very low energy, small radius jets can only be reliably fully calibrated for $\gtrsim 15$ GeV~\cite{jescms,jesatlas}, where the actual threshold may depend on $r$.  This minimum $p_T$ threshold acts as an effective grooming for the re-clustered jets (RC).  This is seen clearly in Fig.~\ref{fig:eventdisplay}, where the blue  large radius jet has many constituents far away from the jet axis (which have low $p_T$) and are not part of the re-clustered jet.   One could choose a more aggressive threshold to, for instance, remove the impact of additional $pp$ collisions (i.e. pileup) on the jets.  A more dynamic grooming scheme, which we call {\it re-clustered jet trimming} (RT), in analogy to large radius jet trimming~\cite{trimming},  sets the $p_T$ cut on the small radius jets based on the large radius jet $p_T$ (calculated before any small-$r$ jets are removed).  Specifically, the RT grooming removes any small radius jet constituent $j$ of a large $R$ re-clustered jet $J$ if $p_T^j<f_\text{cut} \times p_T^J$.  The parameter $f_\text{cut}$ can be optimized for a particular kinematic selection and event topology.  Other grooming schemes are possible, but beyond our scope\footnote{Jet grooming procedures applied to jets-as-inputs have been studied in the past (see for instance Ref.~\cite{Gouzevitch:2013qca}); these and other algorithms can be adopted to the re-clustering paradigm.}.

Due to the increased catchment area of large radius jets over small radius jets, they are more susceptible to contributions from pileup.  Just as there are pileup correction techniques for large radius jets and their subjets, one can benefit from pileup corrections to the small radius jet inputs that propagate to re-clustered jets.  In particular, one can remove  jets from pileup interactions with techniques like JVT~\cite{jvt} or pileup jet identification~\cite{jetid} and can correct the remaining jets with methods like the four-vector jet areas subtraction.   

In the growing field of jet substructure, there are many jet observables which depend explicitly on the jet constituents, not just the jet four-vector.  These techniques are still applicable for re-clustered jets.  Section~\ref{sec:analysis:substructure} discusses two approaches to jet substructure in the re-clustering paradigm.  In a {\it top-down} approach, large radius re-clustered jets inherit the constituents of the small radius jets clustered within.  Clearly, any constituents that might be part of  large radius jets that are not clustered within a small radius jets are not considered under this scheme.  However, this removal of radiation also impacts trimmed  large radius jets.  More details on substructure for trimmed and re-clustered trimmed jets is presented in Section~\ref{sec:analysis:nsub}.  An alternative {\it bottom-up} approach to jet substructure is to use the radius $r$ jets directly as the inputs to jet substructure.   The advantages and limitations of bottom-up substructure are described in Section~\ref{sec:analysis:bottomup}.

\clearpage
\newpage

\section{Simulation}\label{sec:simulation}

Three processes are generated using \textsc{Pythia} 8.170~\cite{Pythia8,Pythia} at $\sqrt{s}=14$ TeV for studying the efficacy of re-clustered jets.  Hadronic $W$ boson and top quarks are used for studying hard 2- and 3-prong type jets.  To simulate high $p_T$ hadronic $W$ decays, $W'$ bosons are generated which decay exclusively into a $W$ and $Z$ boson which subsequently decay in quarks and leptons, respectively.  The $p_T$ scale of the hadronically decaying $W$ is set by the mass of the $W'$ which is tuned to 800 GeV for this study so that the $p_T^W \lesssim 400$ GeV. In this $p_T^W$ range, not all of the decay products of the $W$ are expected to merge into a small radius jet of $r\lesssim 0.4$, but should merge within a cone of $R=1.0$.  A sample enriched in 3-prong type jets is generated with $Z'\rightarrow t\bar{t}$, where the $Z'$ mass sets the energy scale of the hadronically decaying top quarks.  In this analysis, we use $m_{Z'}=1.0$ TeV, which sets $p_T^t \gtrsim 350$ GeV.  To study the impact on signal versus background for re-clustering, QCD dijets are generated with a range of $\hat{p}_T$ that is approximately in the same range as the relevant signal process. Pileup is generated by overlaying additional (between 20 and 80) independently generated minimum-bias interactions with each signal event (denoted NPV = number of primary vertices).

Jet are re-clustered using \textsc{FastJet}~\cite{fastjet} 3.0.3. While the large radius jets can be defined using any set of parameters, the studies in Sec.~\ref{sec:analysis} will use a fixed large jet algorithm: anti-$k_t$ algorithm with $R=1.0$. 
Trimmed jet are constructed as follows: the constituents of the large anti-$k_{t}$ $R=1.0$ jets are used to reconstruct $k_t$ subjets 
with a distance parameter $R_\text{sub}=0.3$ and imposing a minimal requirement on the ratio of the subjet $p_{T}$ divided by the large-$R$ jet $p_{T}$ $(f_{cut})$ of $0.1$. Here, both transverse momenta are pileup corrected using the median-area method described in Ref.~\cite{areas}. The value of $f_{cut}$ chosen yields approximately optimal resolution of the trimmed jet mass for the studied $W'$ sample -- largely independently of the level of pileup considered.  
Note that this is slightly non-standard: in experimental studies, the $f_\text{cut}$ parameter is usually imposed on the uncorrected jet $p_T$, which would require a pileup dependent $f_\text{cut}$ criterion to optimize the mass resolution.

Re-clustering is investigated with a series of schemes for the small radius jets:

\begin{itemize}
	\item {\bf Algorithms} anti-$k_t$, $k_t$, and C/A
	\item {\bf Radius parameters} $r=0.2,0.3$ and $0.4$.
	\item {\bf Grooming} jet $p_T$ cut of 15 GeV for RC and RT with trimming $f_\text{cut}$ parameters $0.1$ and $0.2$.
\end{itemize}

\noindent This list is not exhaustive, but encompasses a relevant set of parameters.   Radii below $r=0.2$ are not considered due to experimental limitations from calorimeter granularity and theoretical considerations from non-trivial non-perturbative effects.  Small radius jets are required to have $p_T>15$ GeV.


\section{Analysis and Results}\label{sec:analysis}

\subsection{Performance of Jet Mass}

\begin{figure}[htbp!]
\begin{center}
\includegraphics[width=0.5\textwidth]{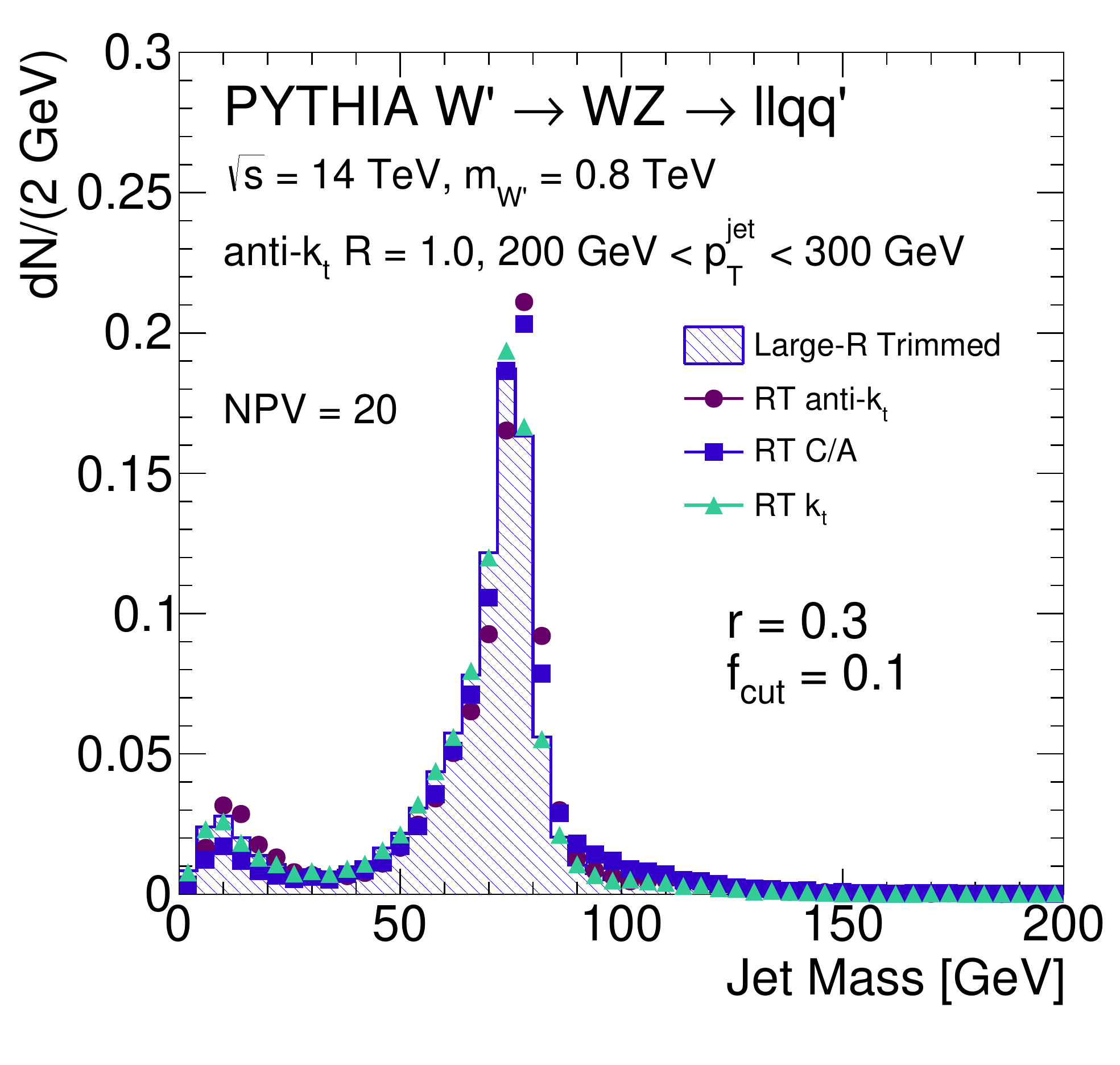}\includegraphics[width=0.5\textwidth]{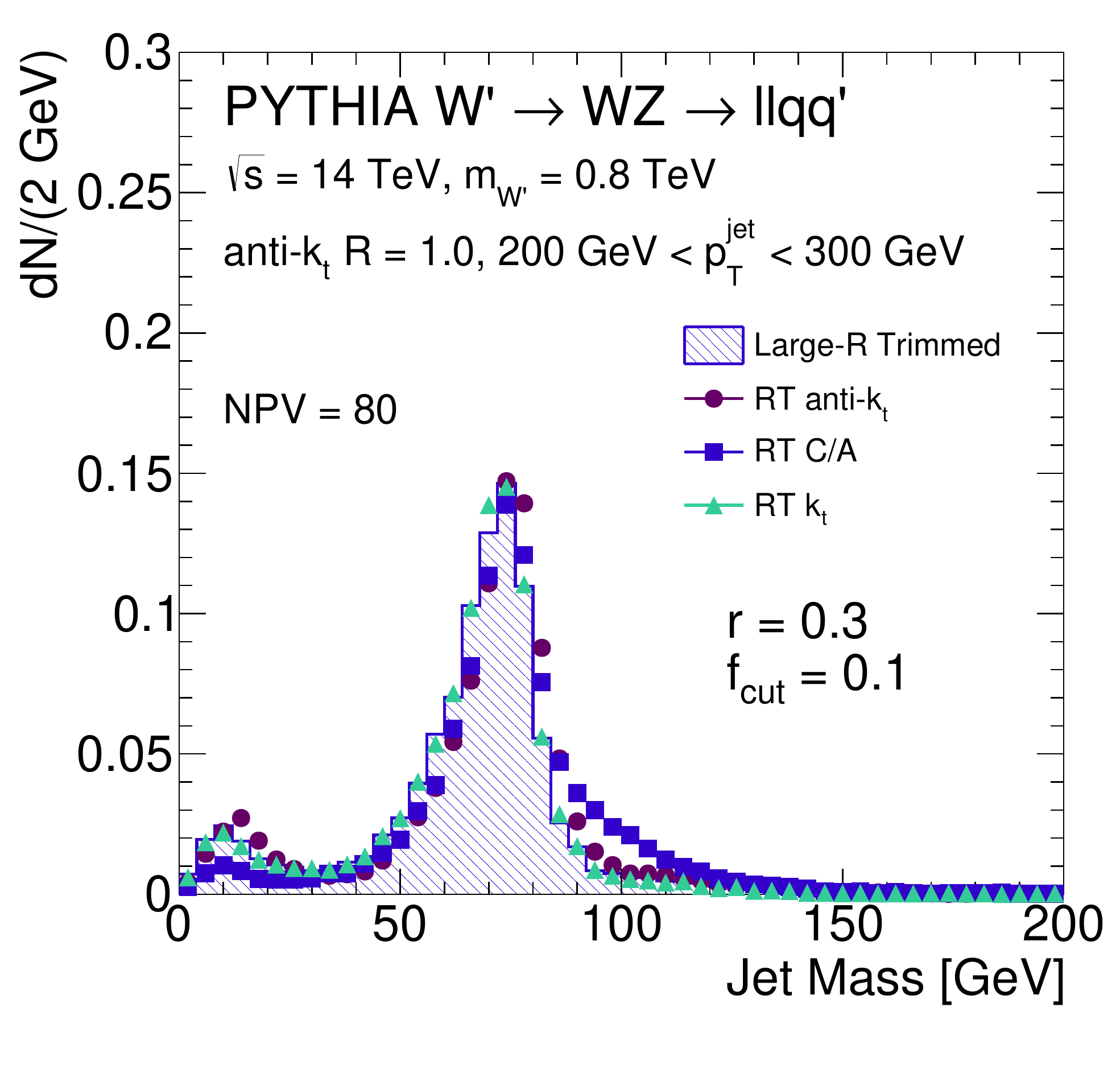}
\end{center}
\caption{Jet mass distribution for various algorithms using small radius jets at fixed $r=0.3$ and $f_\text{cut}=0.1$ for NPV = 20 on the left and NPV = 80 on the right, using a sample of $W^{\prime}\to WZ$ events.}
\label{fig:algo}
\end{figure}

\begin{figure}[htbp!]
\begin{center}
\includegraphics[width=0.5\textwidth]{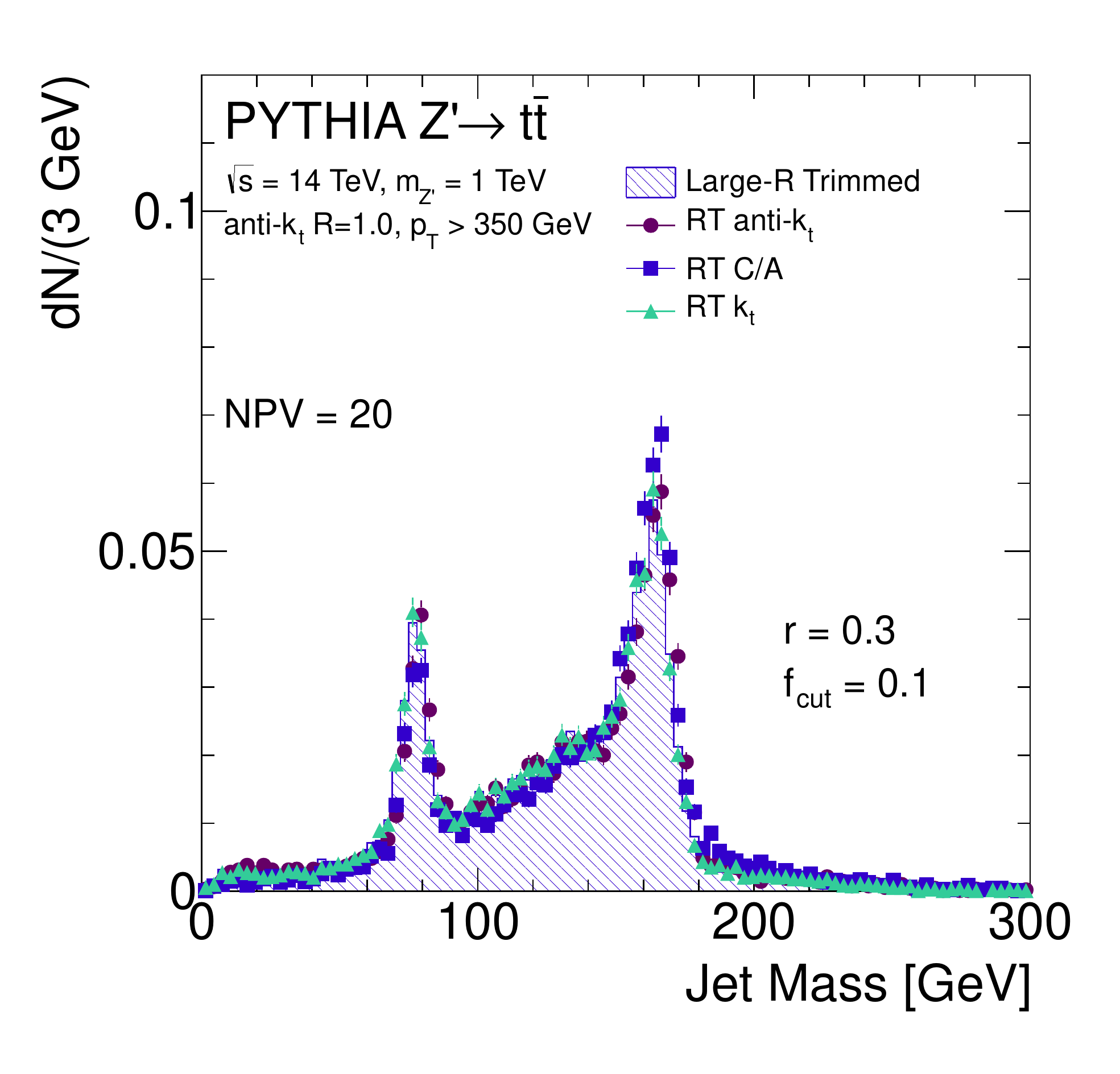}\includegraphics[width=0.5\textwidth]{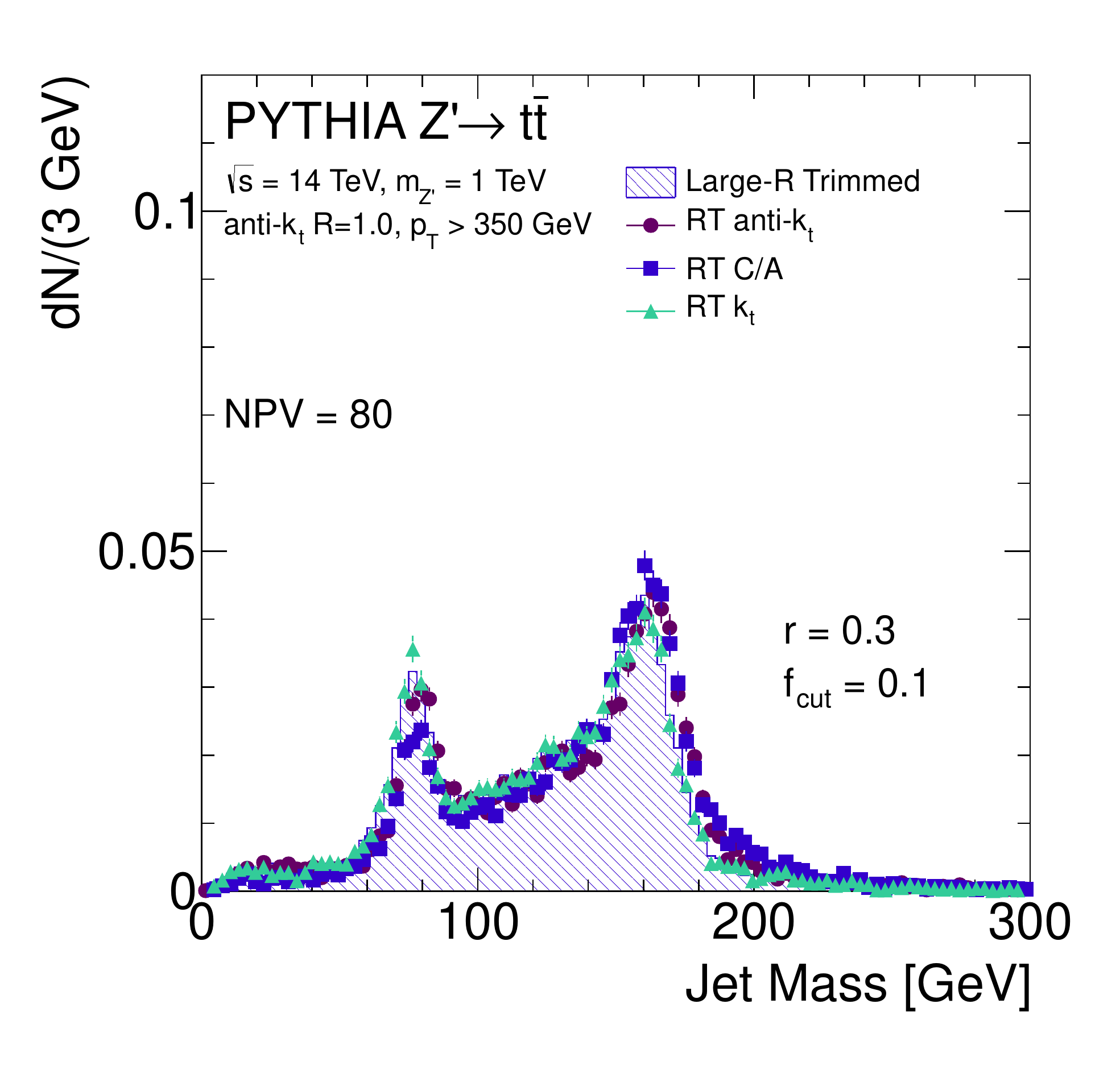}
\end{center}
\caption{Same as Figure~\ref{fig:algo} but using a sample of  $Z^{\prime}\to t\bar{t}$ events. }
\label{fig:algoZprime}
\end{figure}

The most widely used large radius jet observable is the jet mass, $(\sum_{i\in\text{jet}} E_i)^2-(\sum_{i\in\text{jet}} \vec{p}_i)^2$, where $i$ runs over the constituents of the jet.  To compare the performance of  large radius jets with re-clustered jets, we study the performance of the jet mass for the various re-clustering schemes described in Section~\ref{sec:simulation}.  Jet mass performance is quantified by the average jet mass $\langle m\rangle$, a mass resolution, $\sigma$, and the dependance of these quantities with the amount of pileup.  The averages and  deviations are computed over a fixed mass range: 60-100 GeV.  We also present the efficiency of a $60<m_\text{jet}/\text{GeV}<100$ mass cut (fraction of events with a mass in this window) as a third figure of merit.  Figure~\ref{fig:algo} shows the RT jet mass distribution for three small radius jet clustering algorithms with fixed $r=0.3,f_\text{cut}=0.1$.  At very low levels of pileup, there are essentially no differences between building jets from the top down (trimming) or bottom up (re-clustered trimming), as is evident from the fact that $r=0.3$ $k_t$ jets re-clustered with the $R=1.0$ anti-$k_t$ algorithm are nearly identical in distribution to the  large anti-$k_t$ $R=1.0$ trimmed jets with $k_t$ subjets.  However, with increasing levels of pileup, there are clear differences between the three algorithms.  For instance, the C/A algorithm has a large high mass tail which is absent in the other algorithms.  However, the tradeoff is that C/A has a smaller low mass peak that arises when some of the small radius jets are removed to effectively reconstruct the $W$ jet. Figure~\ref{fig:algoZprime} is similar to Figure~\ref{fig:algo} but uses a sample of $Z^{\prime}\to t\bar{t}$ events. Again, especially for low pileup levels, the difference between  trimming and the different re-clustering configurations are small. The optimal re-clustering configuration, however, is topology dependent and not necessarily the same for the considered $W^{\prime}$ and $Z^{\prime}$ samples. In the following, we will focus our studies on the $W^{\prime}$ events.

\begin{figure}[htbp!]
\begin{center}
\includegraphics[width=0.5\textwidth]{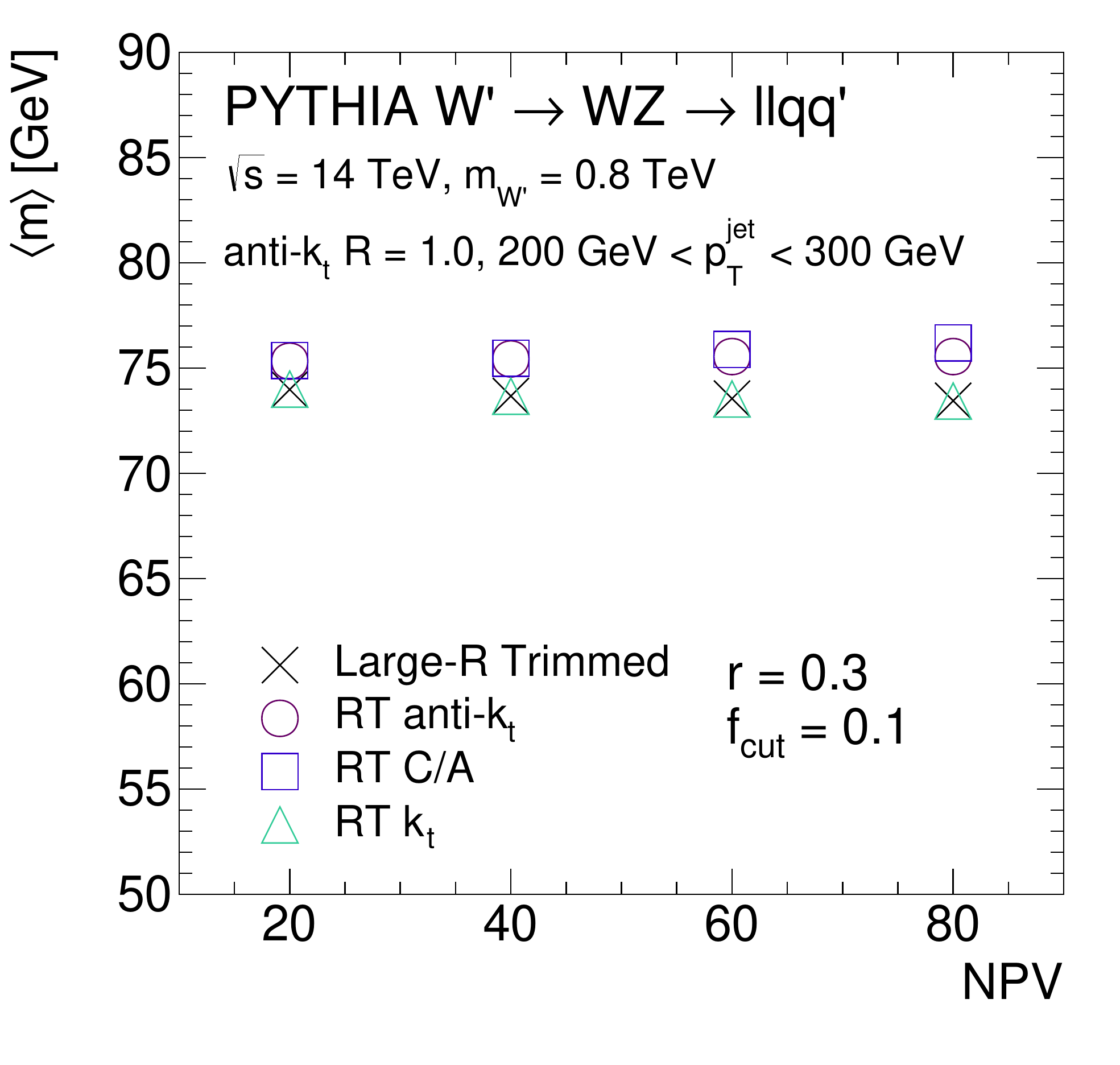}\includegraphics[width=0.5\textwidth]{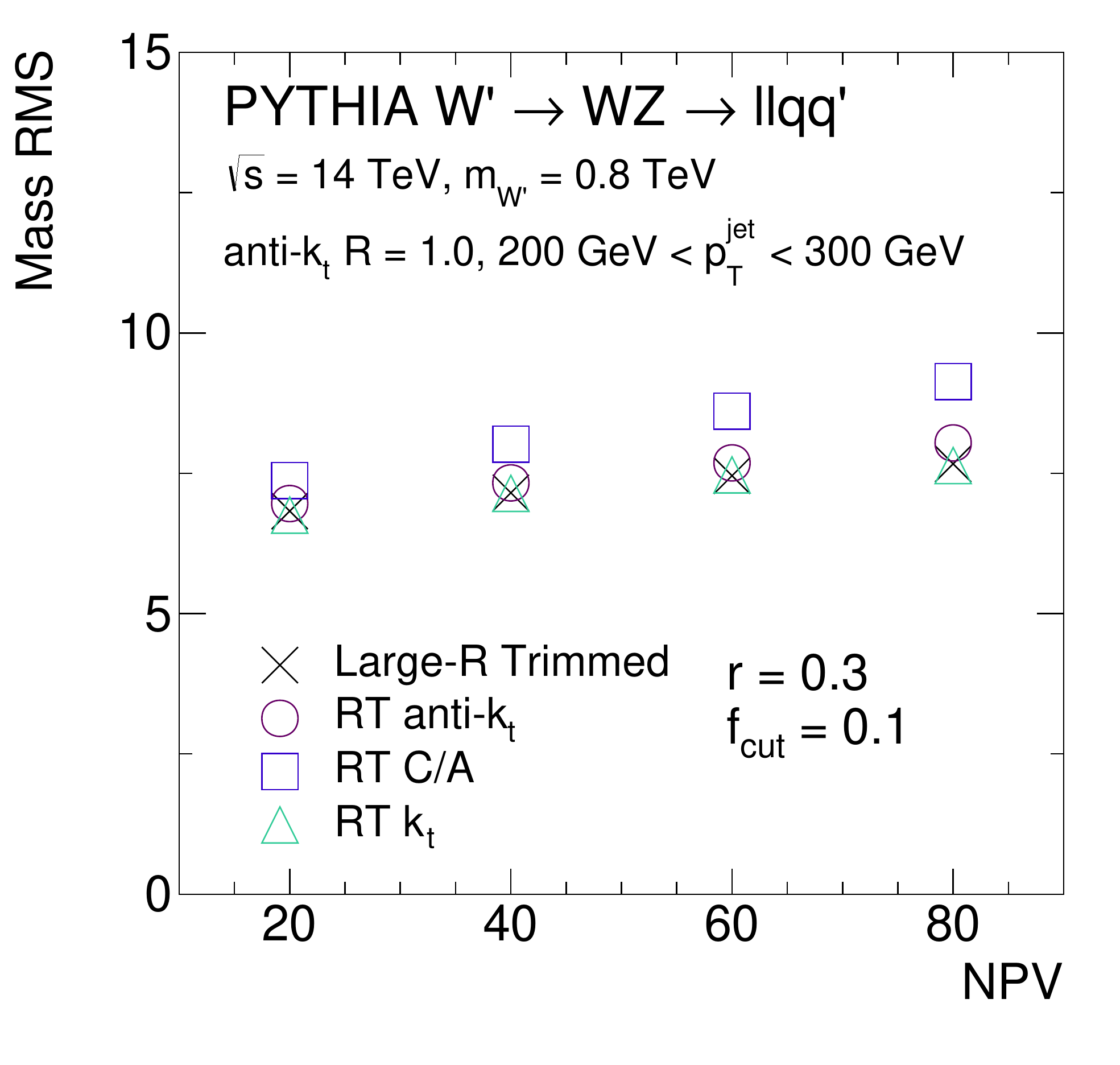}
\includegraphics[width=0.5\textwidth]{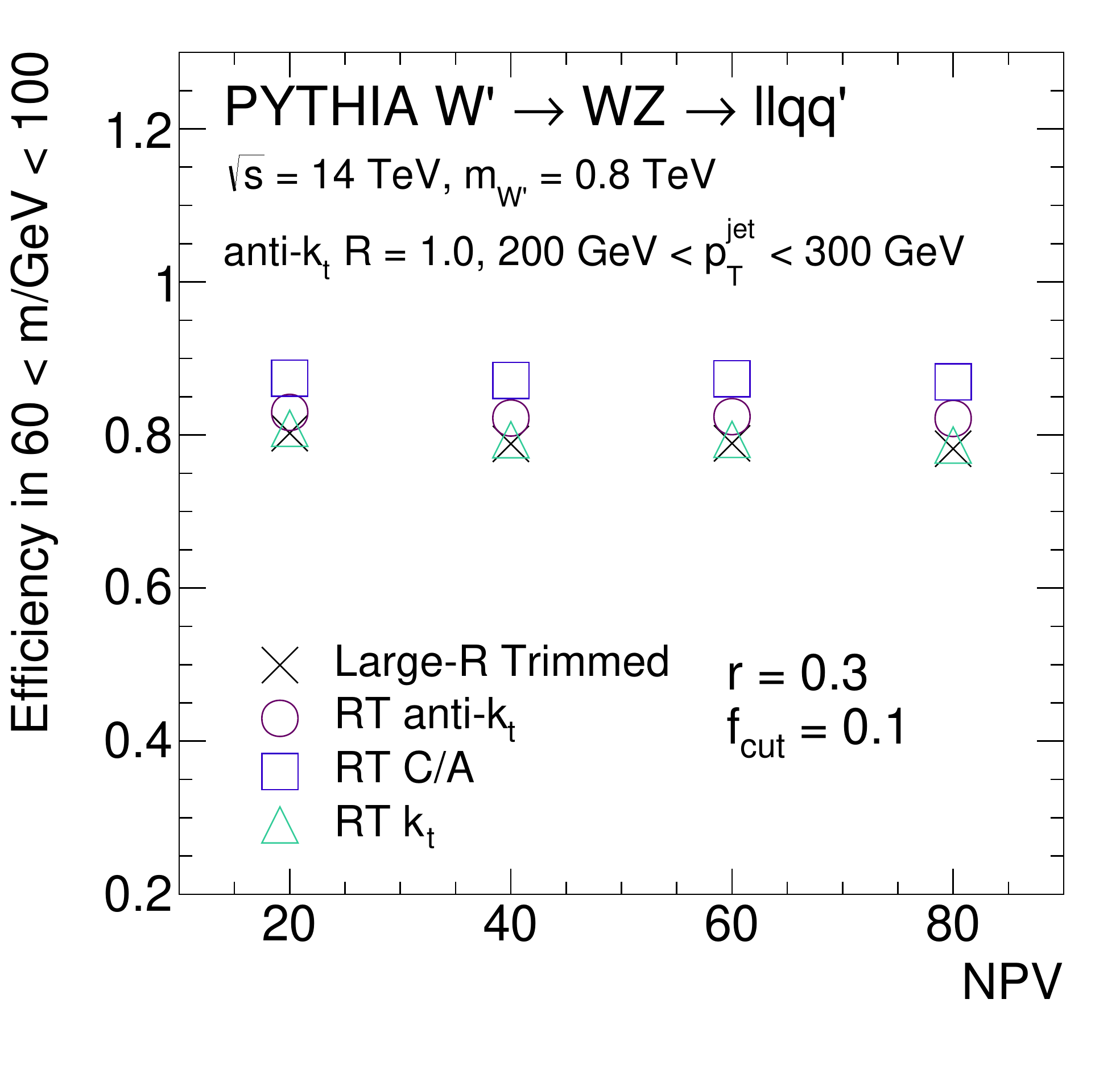}
\end{center}
\caption{Mean, mass resolution, and mass window efficiency of the mass distribution as a function of the number of additional interaction vertices for various small radius jet clustering algorithms.}
\label{fig:algo2}
\end{figure}

The discussed features of Fig.~\ref{fig:algo} are quantified in Fig.~\ref{fig:algo2}. It is seen that the average mass is higher for C/A jets compared to anti-$k_t$ or $k_t$, but the efficiency is also highest for C/A jets. Both of these metrics remain stable under increased levels of pileup.

\begin{figure}[htbp!]
\begin{center}
\includegraphics[width=0.5\textwidth]{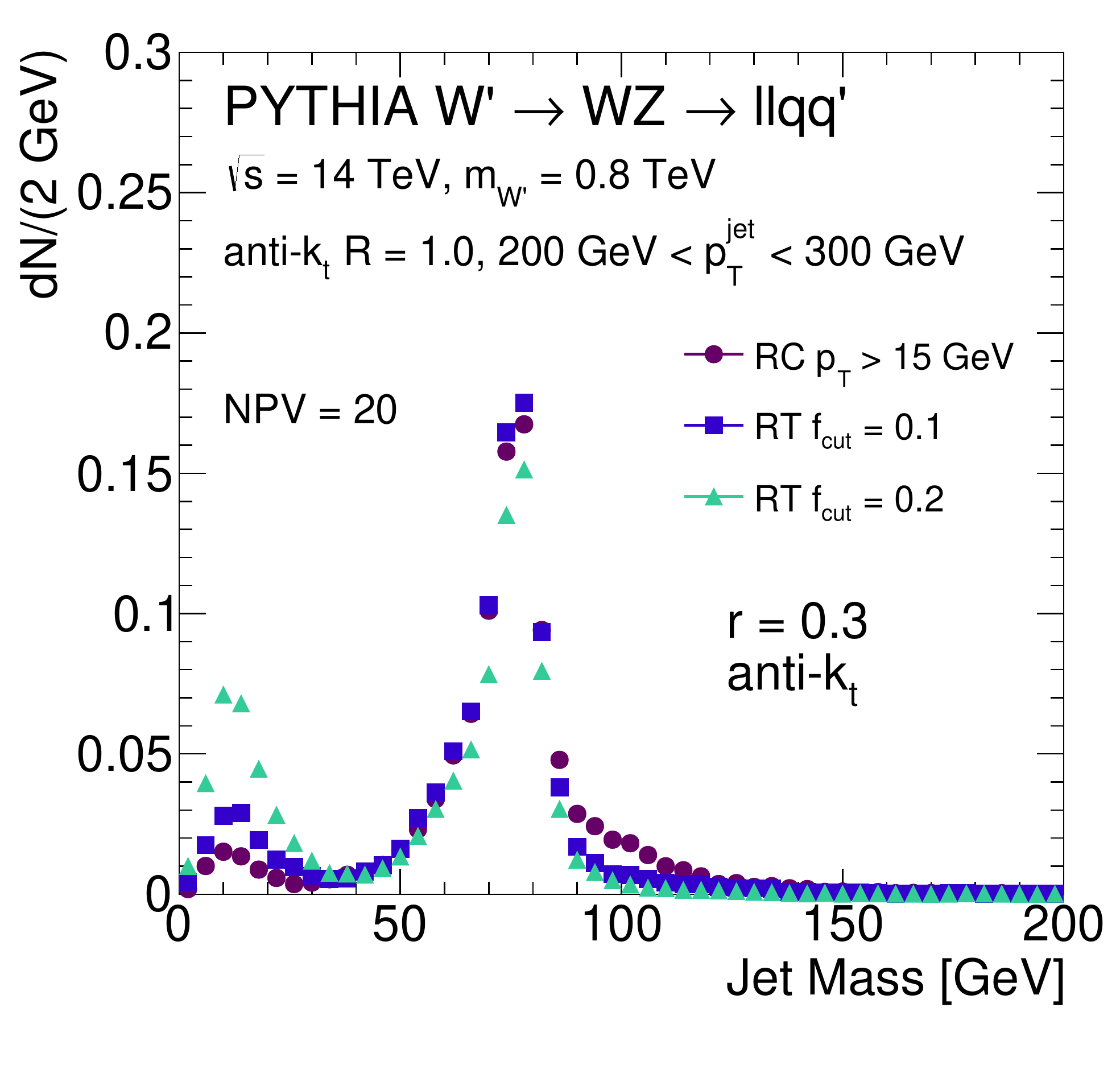}\includegraphics[width=0.5\textwidth]{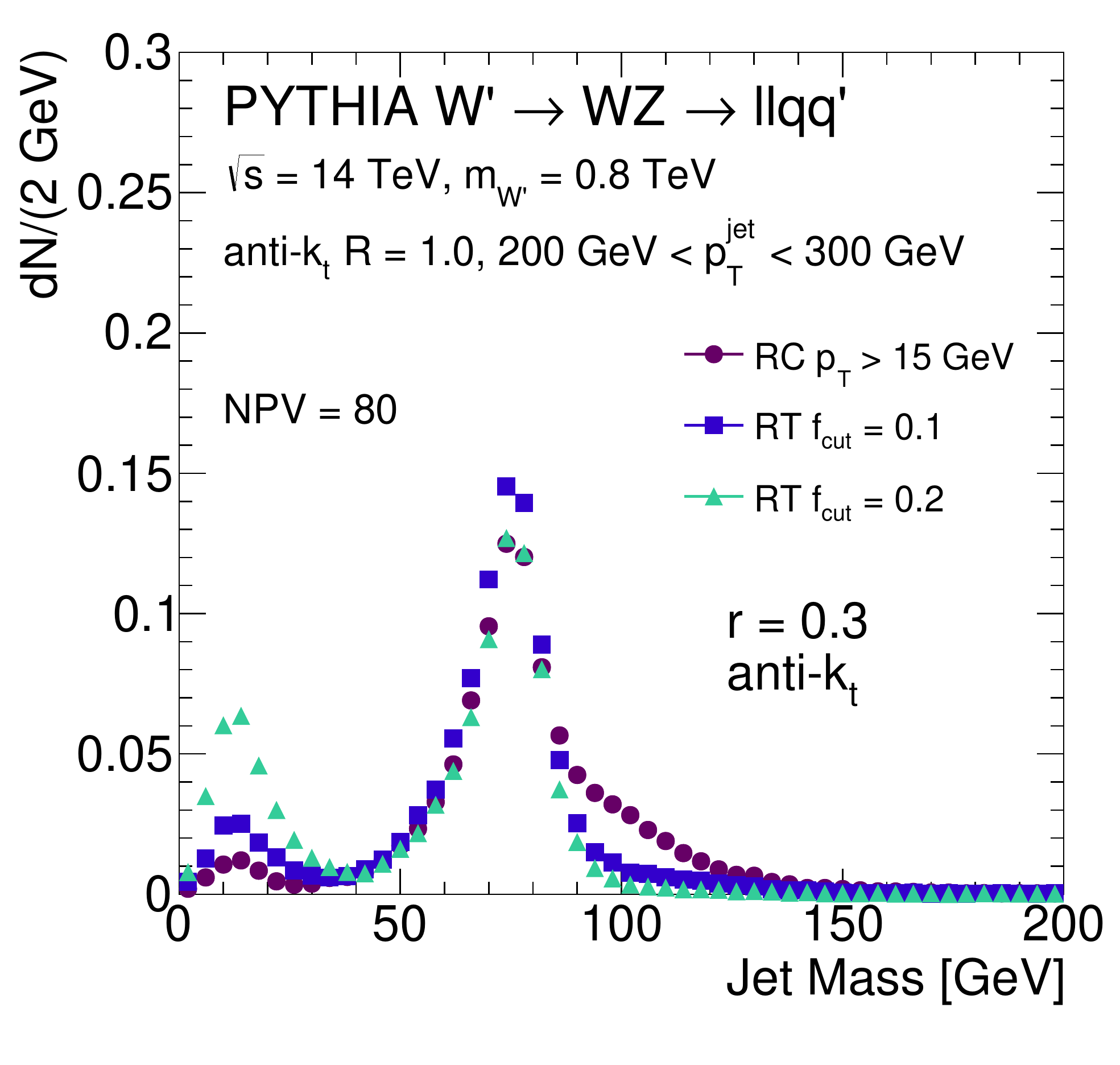}
\end{center}
\caption{Various re-clustered grooming parameters for anti-$k_t$ $r=0.3$ jets for NPV = 20 on the left and NPV = 80 on the right.}
\label{fig:fixed}
\end{figure}

\begin{figure}[htbp!]
\begin{center}
\includegraphics[width=0.5\textwidth]{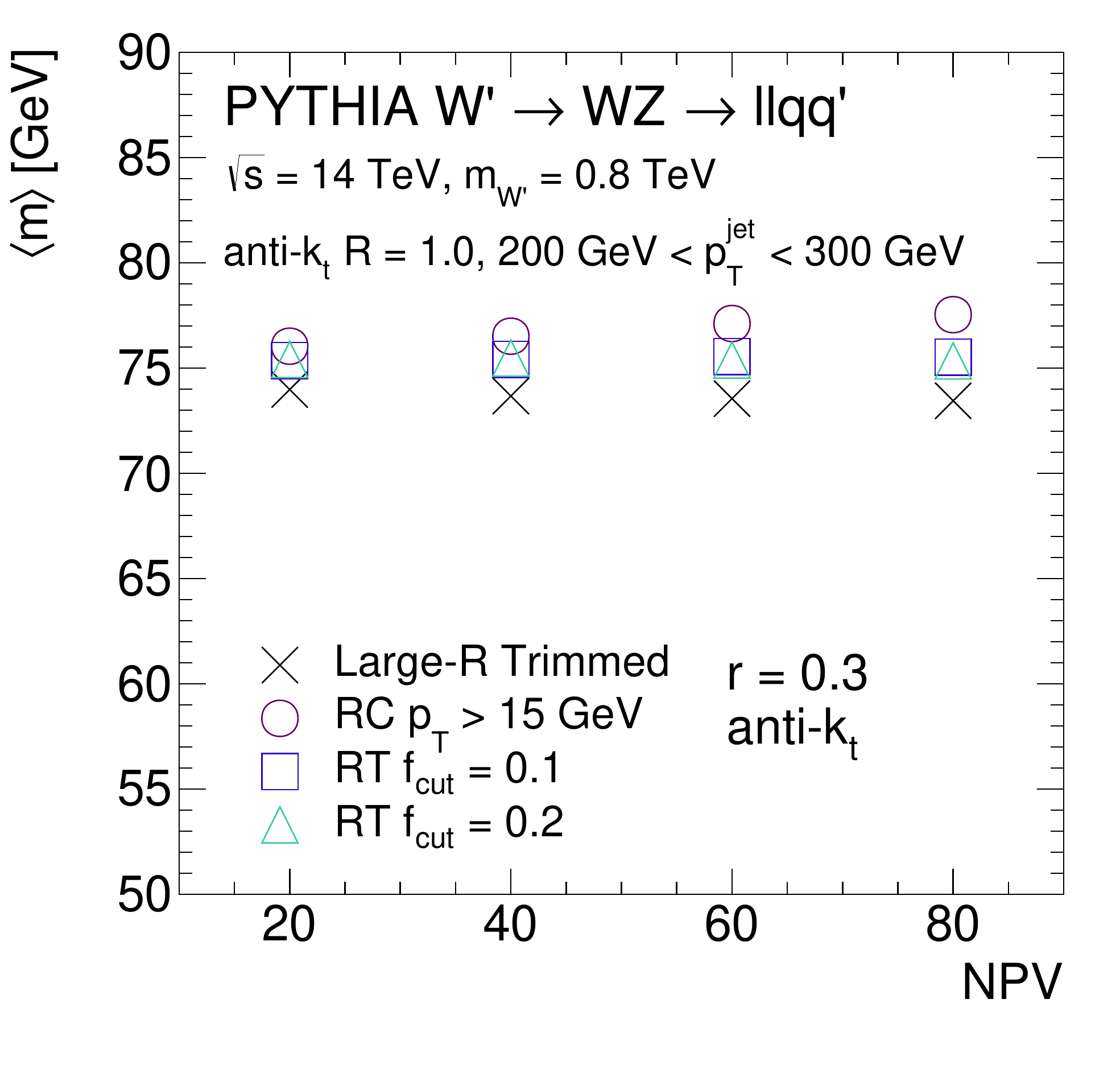}\includegraphics[width=0.5\textwidth]{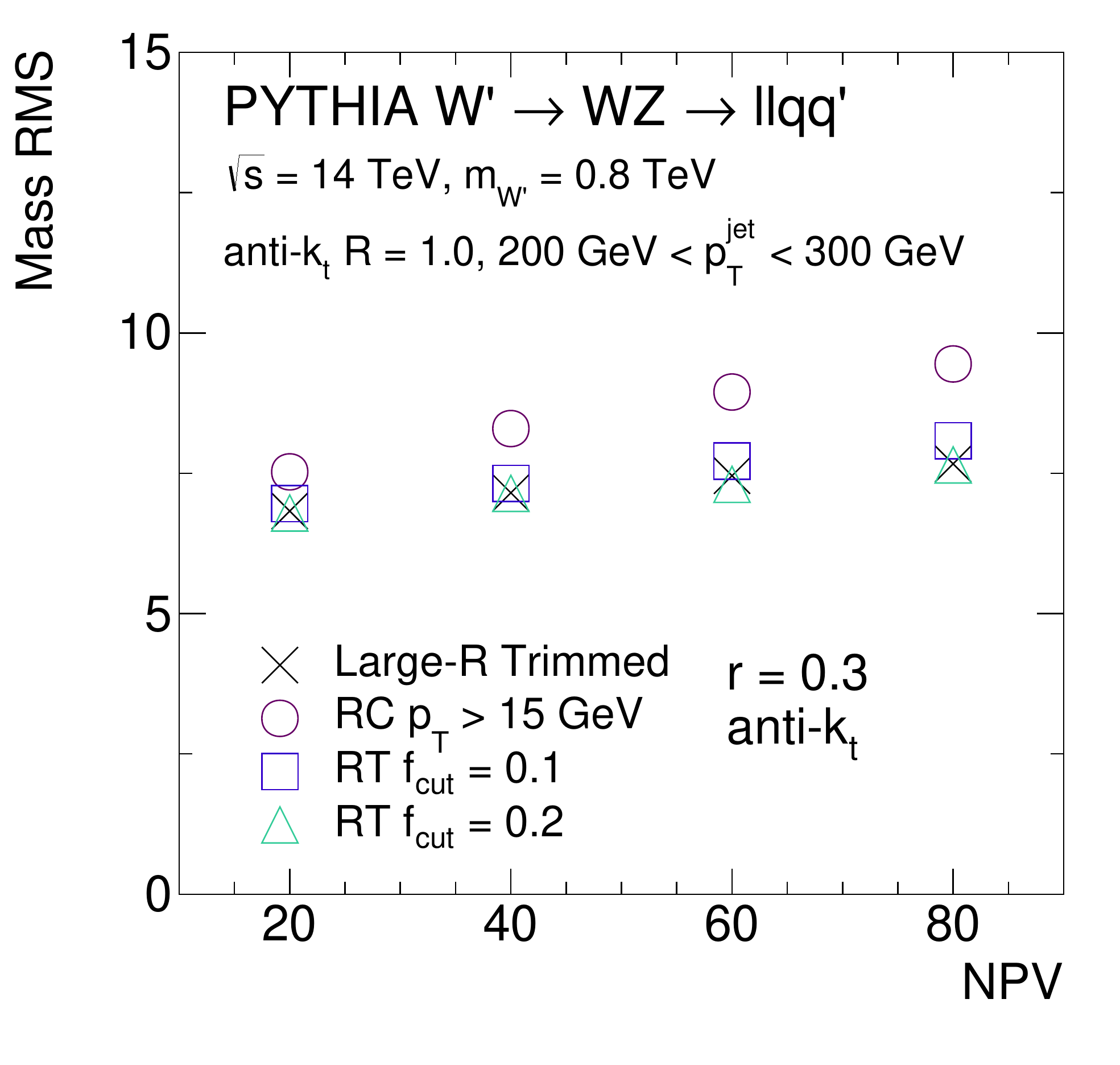}
\includegraphics[width=0.5\textwidth]{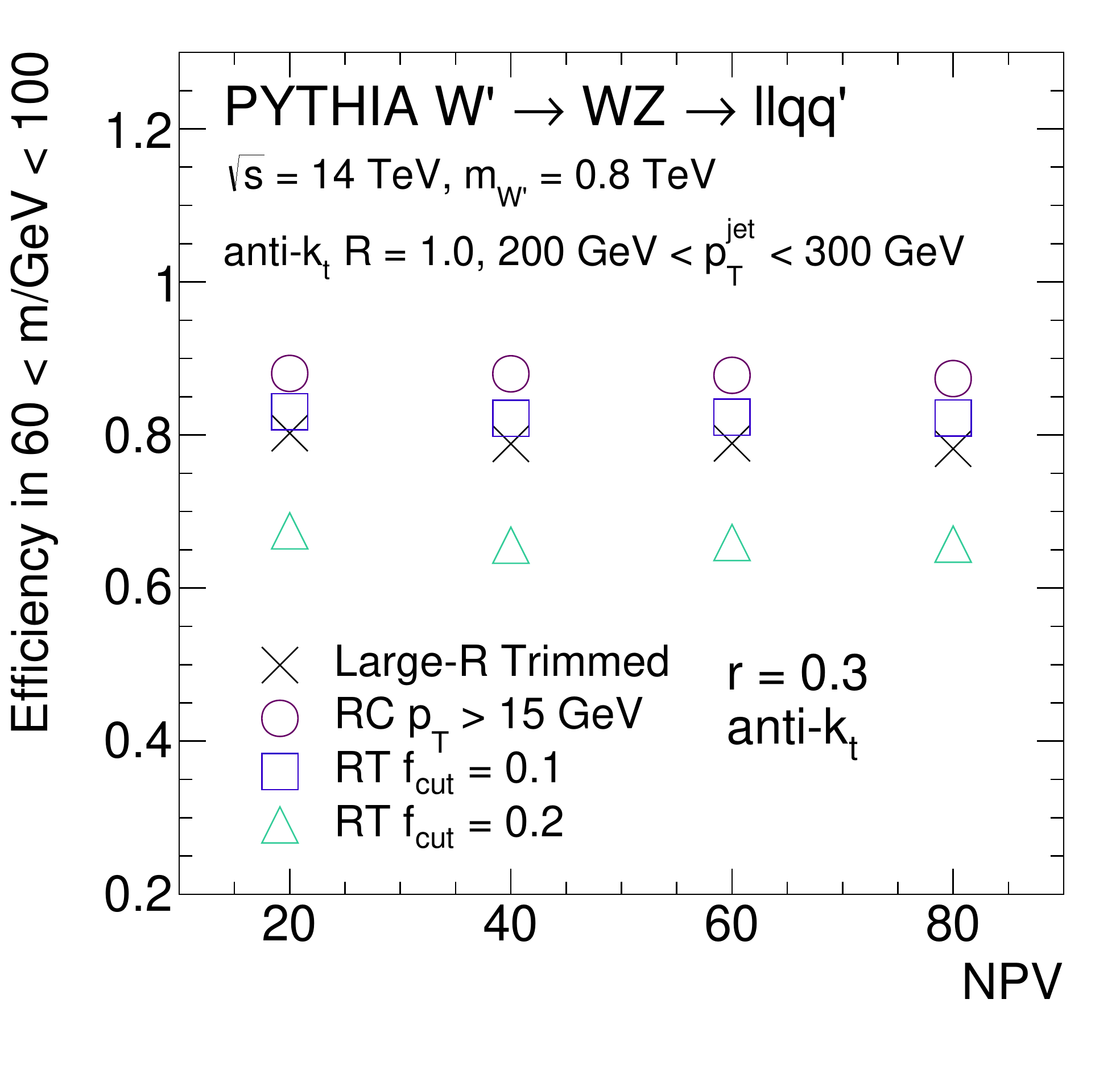}
\end{center}
\caption{Mean, mass resolution, and mass window efficiency of the mass distribution as a function of the number of additional vertices for various re-clustered jet grooming schemes.}
\label{fig:fixed2}
\end{figure}

Figures~\ref{fig:fixed} and~\ref{fig:fixed2} compare trimming to several grooming schemes for re-clustered jets.  In the region near the $W$ mass peak, re-clustered trimming with $f_\text{cut}=0.2$ performs the best.  However, there is a sizable peak at low mass where too many jets have been cut out by the aggressive trimming parameter.  The fixed cut of $15$ GeV is too low, especially at very high pileup where the large high mass tail is much bigger for RC than for RT.  The re-clustered trimming using anti-$k_t$ with the same $f_\text{cut}$ as the  trimming has very similar performance, though the peak position is slightly higher.  Figure~\ref{fig:fixed2} shows the performance metrics as a function of NPV for the various grooming schemes.  The average mass for RT is very stable, whereas there is a slight slope for RC.   The mass resolution for RC is slightly worse than for RT, but the efficiency of RC is better because it avoids the peak at low masses well below the $W$ boson mass.

\begin{figure}[htbp!]
\begin{center}
\includegraphics[width=0.5\textwidth]{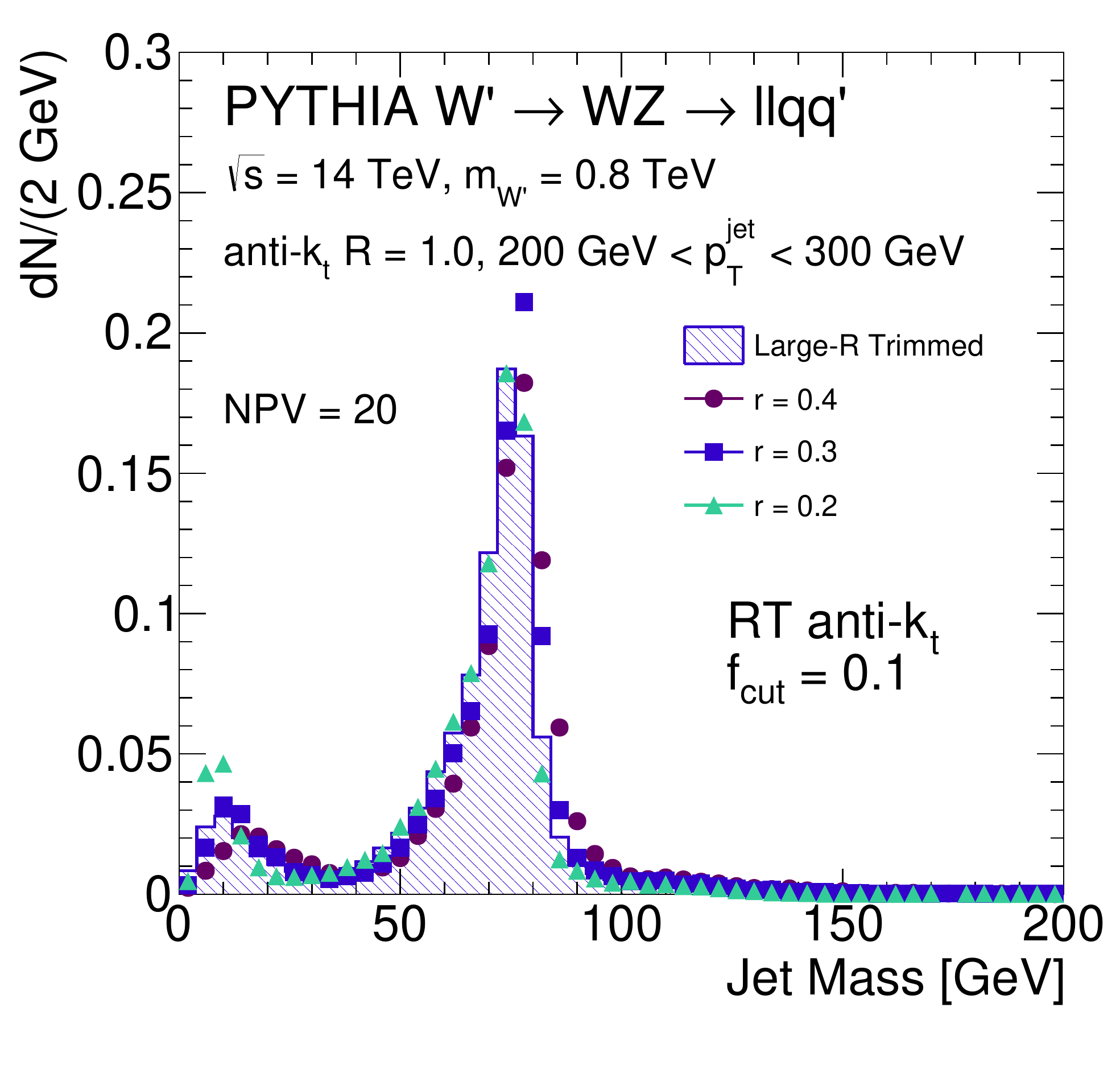}\includegraphics[width=0.5\textwidth]{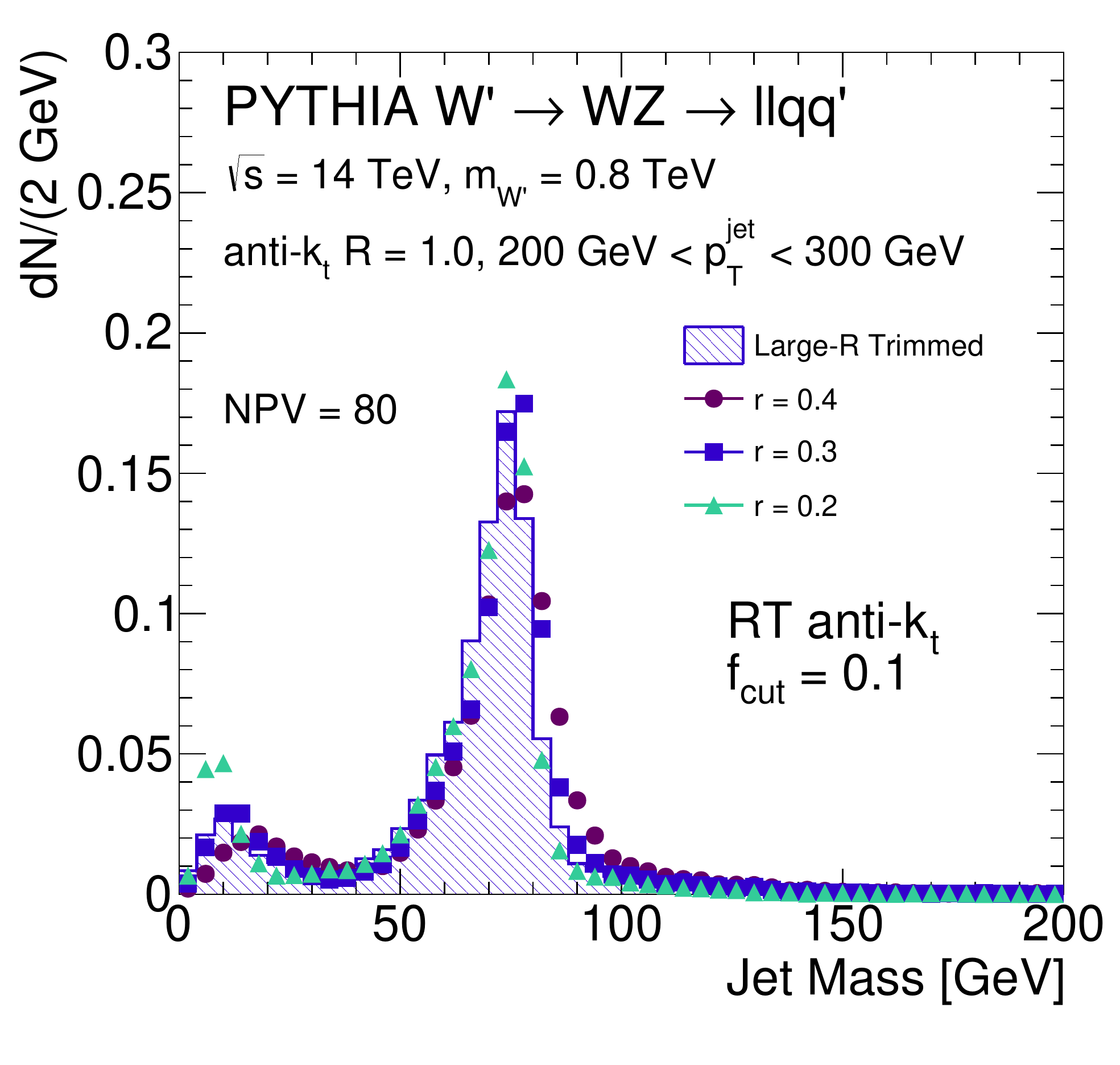}
\end{center}
\caption{Various small radii for a fixed algorithm of anti-$k_t$ for NPV = 20 on the left and NPV = 80 on the right.}
\label{fig:sizes}
\end{figure}

\begin{figure}[htbp!]
\begin{center}
\includegraphics[width=0.5\textwidth]{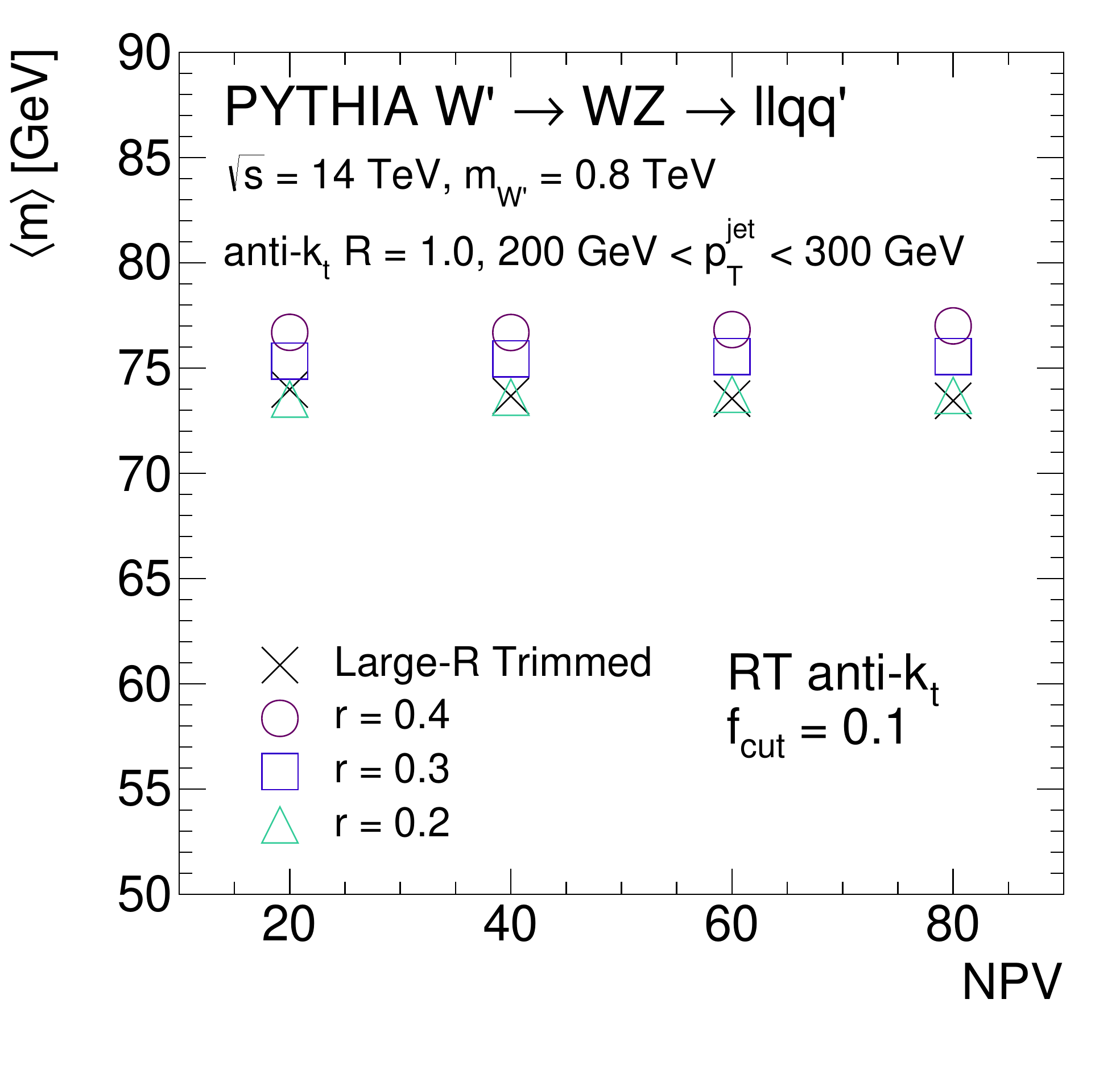}\includegraphics[width=0.5\textwidth]{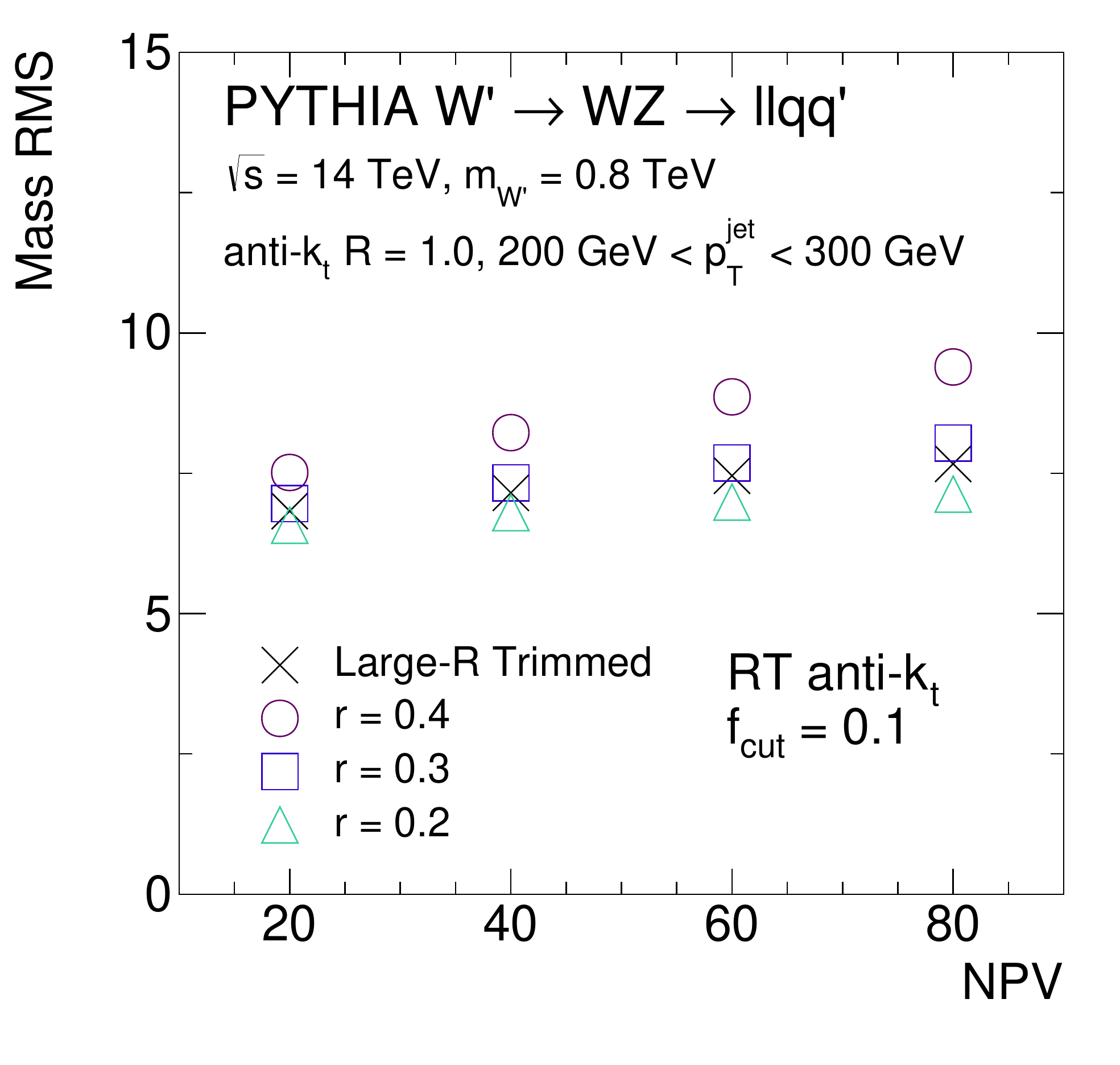}
\includegraphics[width=0.5\textwidth]{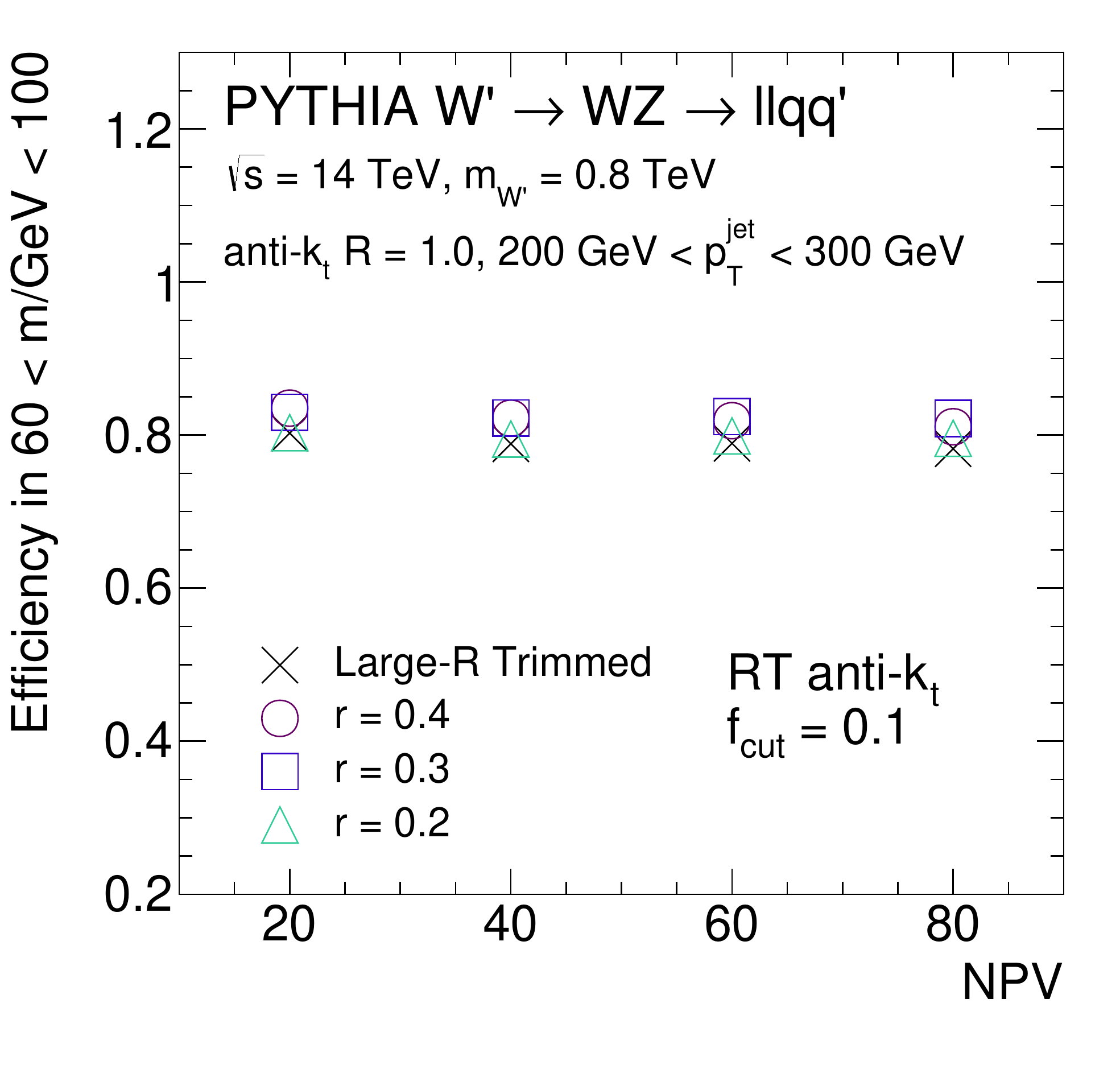}
\end{center}
\caption{Mean, mass resolution, and mass window efficiency of the mass distribution as a function of the number of additional vertices for various small jet radii.}
\label{fig:sizes2}
\end{figure}

The mass distribution for several small radius jet sizes is shown in Figure~\ref{fig:sizes} and the performance metrics are quantified in Fig.~\ref{fig:sizes2}.  For all three considered values of $r$, the minimum $p_T$ cut is 15 GeV.  In practice, this could be optimized, since smaller radius jets may be calibrated at smaller values of $p_T$.  An alternative approach is to use {\it iterative re-clustering} by re-clustering $r=0.2$ into $r'=0.4$ and then into $R=1.0$ to further increase the flexibility of the jet algorithms (also this reduces the effective jet area and so the resulting jets would be less susceptible to pileup\footnote{If viewed as a uniform noise in the calorimeter, the contribution of pileup to a given jet scales proportionally to its area.  There are, however, local fluctuations that complicate this picture.}).  The right plot of Figure~\ref{fig:sizes} and the top right plot of Fig.~\ref{fig:sizes2} show the $r=0.2$ as the most peaked.  

\clearpage
\newpage

\subsection{Comparisons to QCD}

\begin{figure}[htbp!]
\begin{center}
\includegraphics[width=0.33\textwidth]{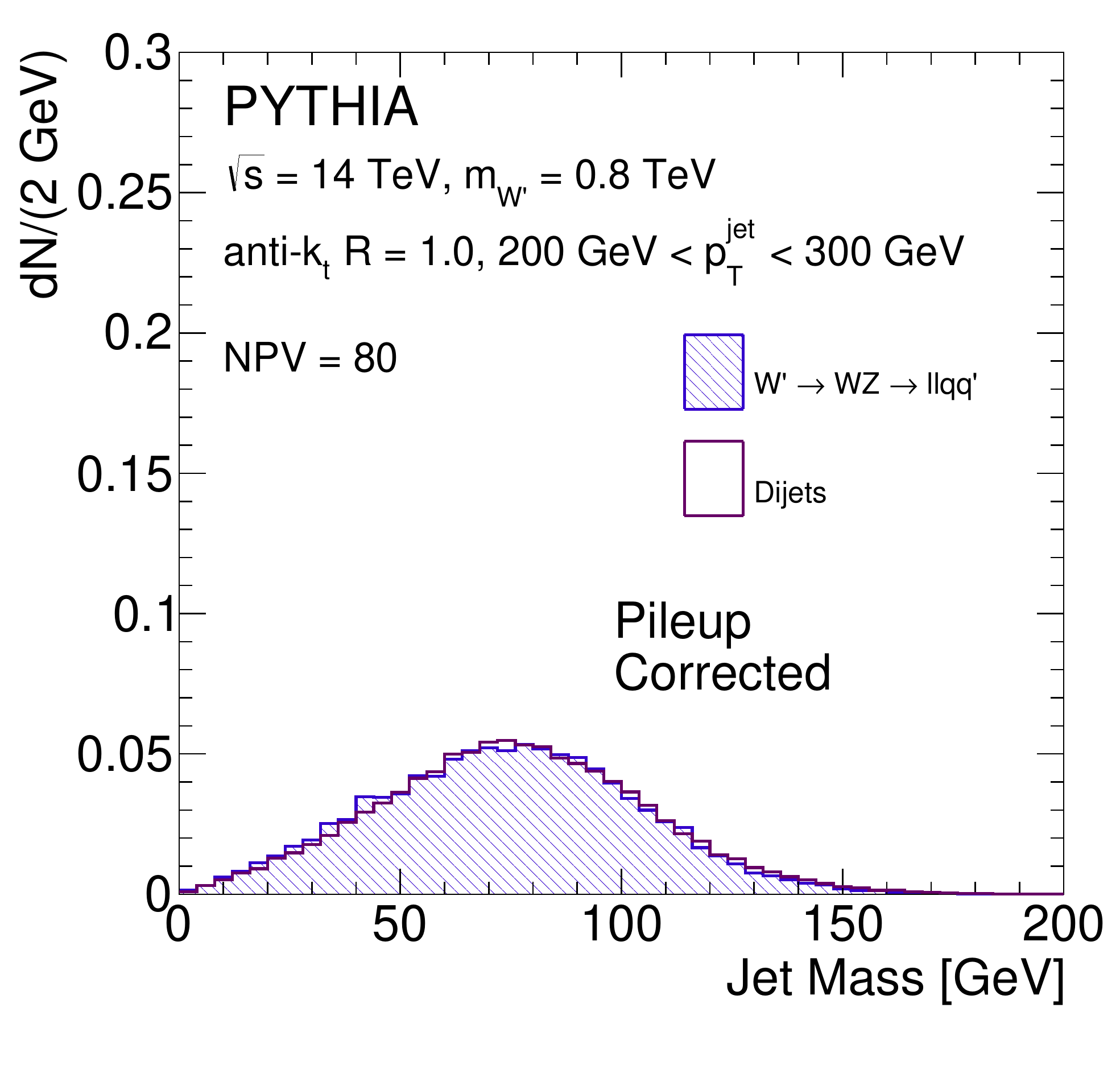}\includegraphics[width=0.33\textwidth]{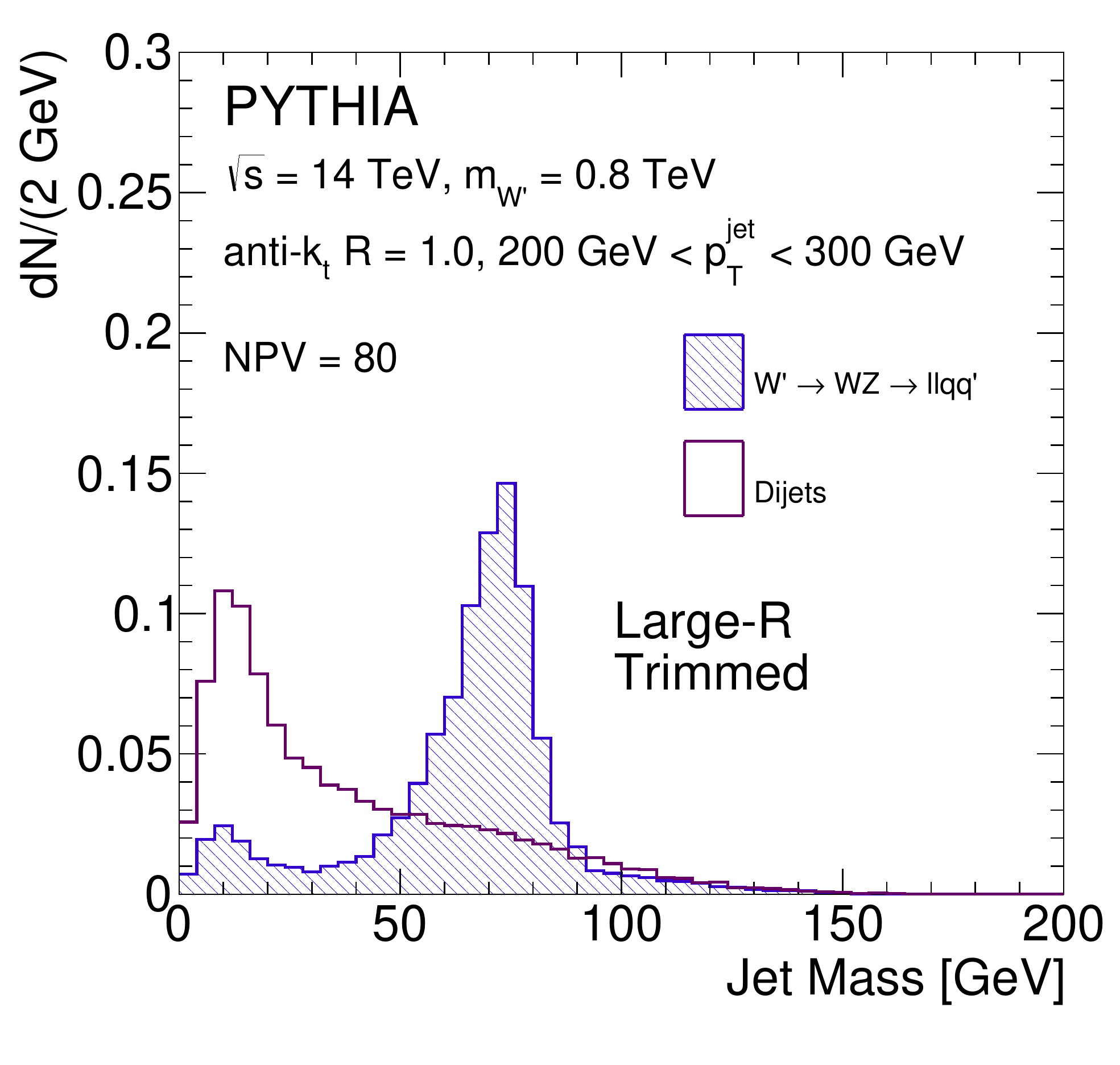}\includegraphics[width=0.33\textwidth]{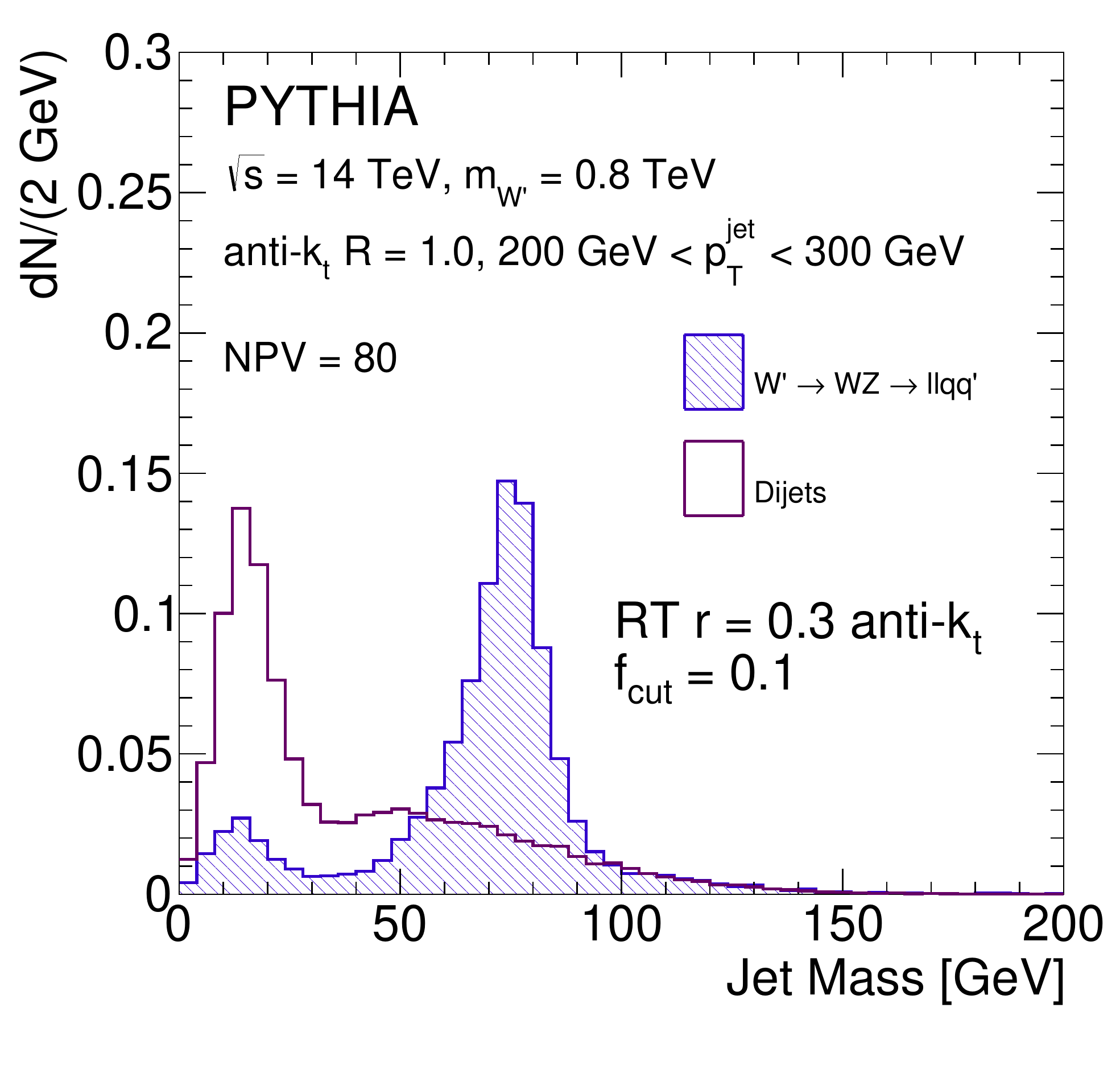}
\end{center}
\caption{Comparison of jet mass between $W$ jets and QCD jets, for 4-vector pileup corrected jets (left), trimmed jets (center) and re-clustered jets (right). Re-clustered jets use anti-k$_t$ $R=0.3$ jets as inputs, and apply a trimming cut of $f_{cut}=10\%$. A pileup level of $NPV=80$ is used.}
\label{fig:m_qcd}
\end{figure}

It is important to study the effect of re-clustering not just on signal jets, but also on the most likely background-- jets originating from QCD multijet processes. It is well established that various jet grooming techniques increase the separation in jet mass between signal and QCD jets \cite{boost2011,boost2012}. Figure~\ref{fig:m_qcd} shows this comparison: while a 4-vector pileup correction alone (left) does not allow for separation between QCD and $W$ jet, both trimming (center) and re-clustering (right) allow for the successful discrimination of signal and background using the jet mass. The details of the optimization to improve the signal-to-background in a mass window are best left up to the experimental analyses which use these techniques, but figure~\ref{fig:m_qcd} indicates that similar performance in QCD rejection is possible with re-clustering as compared to the use of large-$R$ trimmed jets.

\section{Re-clustering and Jet Substructure}
\label{sec:analysis:substructure}

Jet substructure techniques have become very sophisticated tools for classifying hadronic final states.  There are two natural ways to define jet substructure within the re-clustering paradigm.  In a {\it top-down} approach, re-clustered jets inherit constituents from the radius $r$ jets (Sec.~\ref{sec:analysis:nsub}).  In a {\it bottom-up} approach, the radius $r$ jets are themselves the inputs to calculating jet substructure properties of the reclustered radius $R$ jet (Sec.~\ref{sec:analysis:bottomup}).

\subsection{Performance of Shape Variables}
\label{sec:analysis:nsub}

In addition to studying the jet mass, in this section we consider one example of a jet substructure variable, the $n$-subjettiness ratio $\tau_{21}$, and compare QCD to $W$ jets \cite{nsub}. $N$-subjettiness moments are defined over a set of $N$ axes \footnote{We use the ``one-pass'' $k_{t}$ axes optimization technique, which uses an exclusive $k_t$ algorithm to find $N$ axes and then refines them by minimizing the n-subjettiness value.}, and calculated as:
\begin{equation}
\tau_N = \frac{1}{d_0} \sum_k p_{T,k} \min \{ \Delta R_{1,k}, \Delta R_{2,k}, \ldots \Delta R_{N,k} \}
\end{equation}
with the normalization is defined as:
\begin{equation}
d_0 = \sum_k p_{T,k} R_0
\end{equation}
where $R_0$ is the radius of the jet. In practice, analyses use the n-subjettiness ratios:
\begin{equation}
\tau_{ij} = \frac{\tau_i}{\tau_j}
\end{equation}
$\tau_{21}$ is often used for the separation of $W$ from QCD jets~\cite{wbosonCMS,wbosonATLAS}. This variable measures the compatibility of jets with a 2-prong hypothesis compared to a 1-prong hypothesis, where a low value indicates that the jet likely has a 2-prong structure. When calculating n-subjettiness for re-clustered jets, we use as inputs the constituents of the small jets which have been re-clustered: i.e., we use the truth particles (or in a detector, the particle flow objects or topological clusters) and not the small jets themselves. A pre-selection on the $p_T$ and mass, of $200 < p_T / $GeV $< 300 $ and $60 < m / $GeV $ < 100 $ is applied to all jets considered. 

\begin{figure}[htbp!]
\begin{center}
\includegraphics[width=0.5\textwidth]{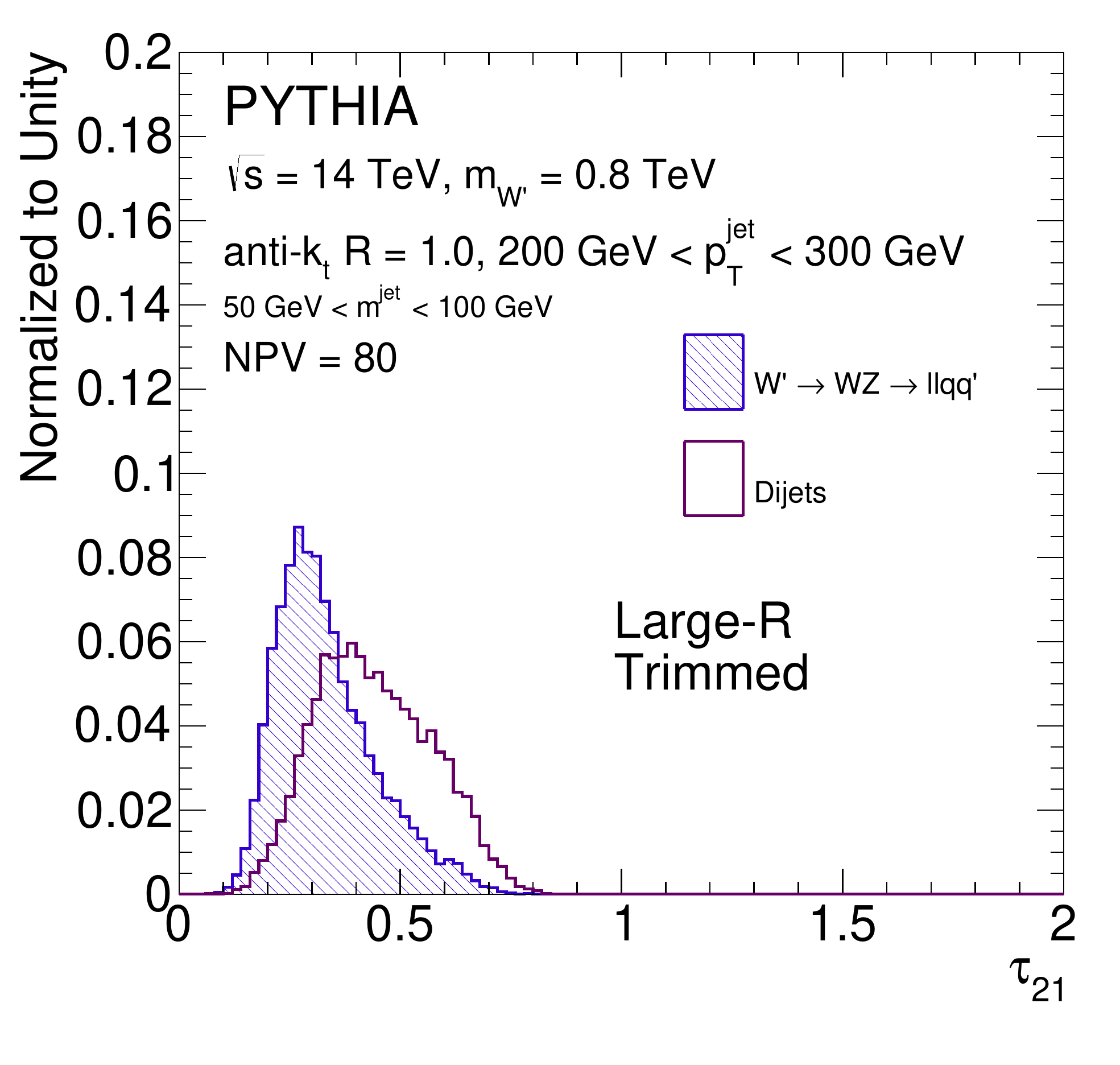}\includegraphics[width=0.5\textwidth]{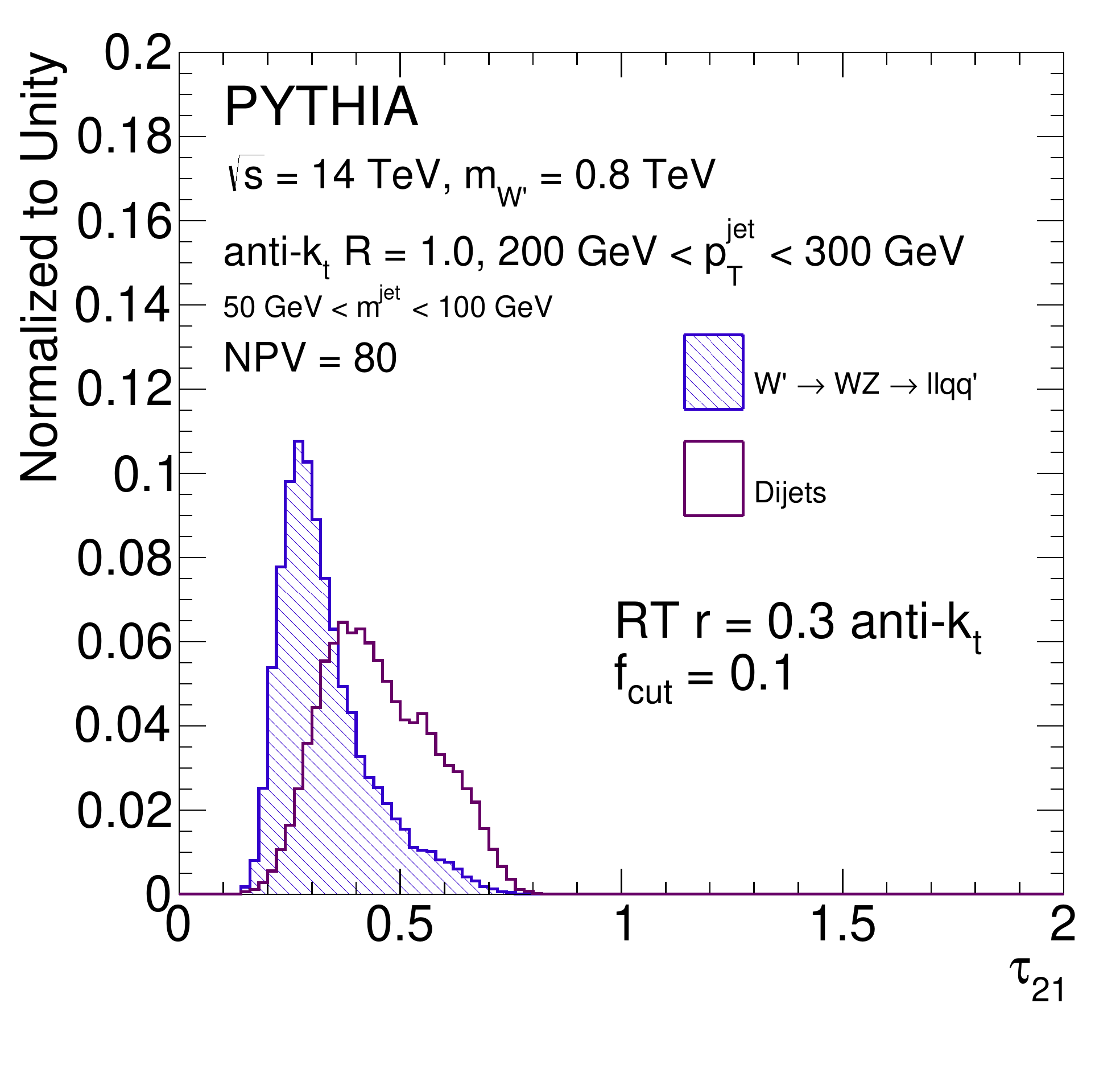}
\end{center}
\caption{Comparison of the n-subjettiness ratio $\tau_{21}$ for $W$ and QCD jets, calculated using trimmed jets (left) and re-clustered jets (right). Re-clustered jets use anti-k$_t$ $R=0.3$ jets as inputs, and apply a trimming cut of $10\%$.  A pileup level of $NPV=80$ is used.}
\label{fig:nsub1}
\end{figure}

\begin{figure}[htbp!]
\begin{center}
\includegraphics[width=0.75\textwidth]{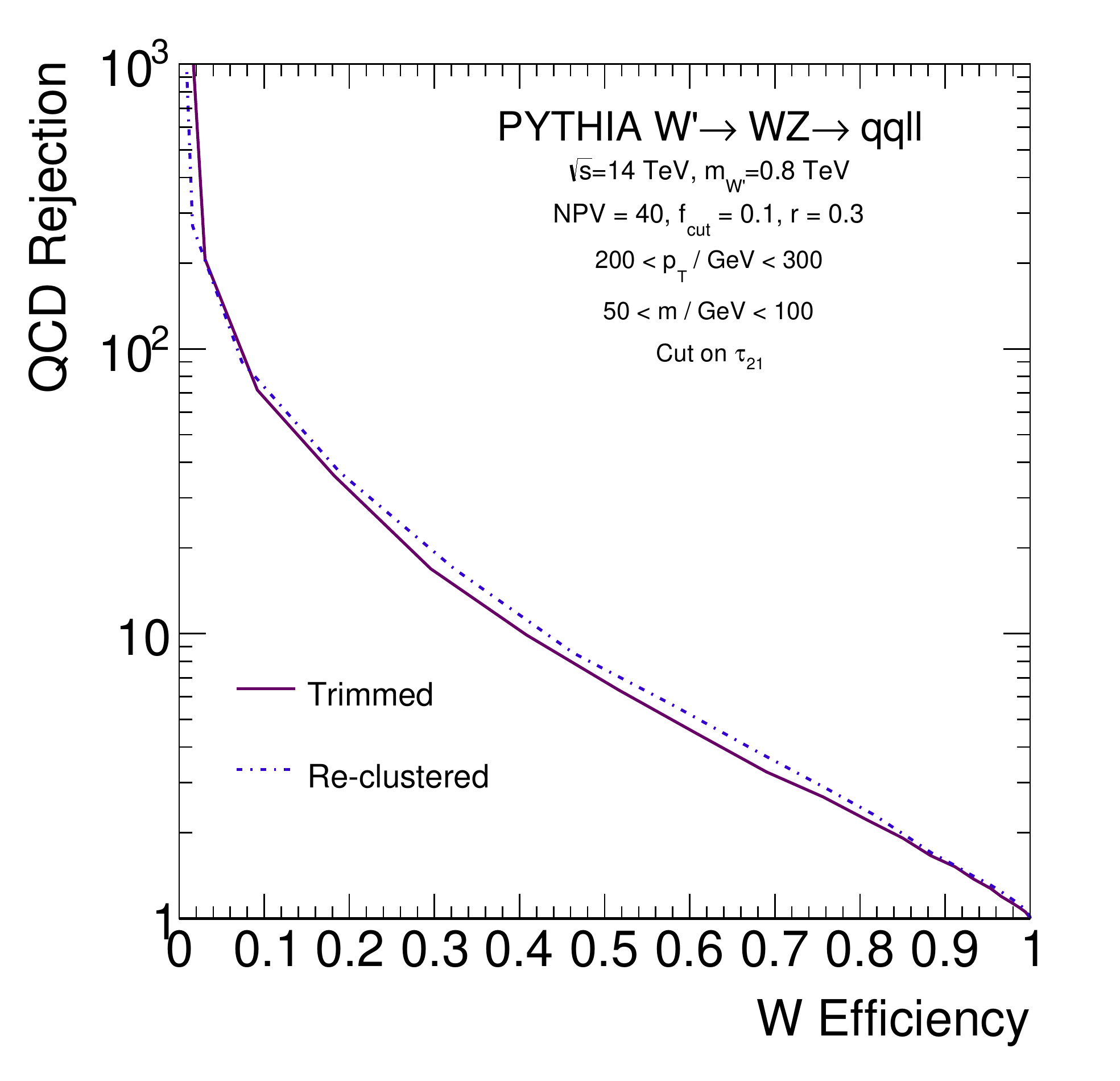}
\end{center}
\caption{Comparison of the efficacy of cuts on n-subjettiness in separating $W$ from QCD jets. Samples with a pileup of NPV = 40 are shown.}
\label{fig:nsub2}
\end{figure}

Figure~\ref{fig:nsub1} shows a comparison of $\tau_{21}$ calculated using trimmed jets (left) and re-clustered jets (right) for samples with $NPV = 80$. Both jet configurations show similar discrimination. For a quantitative comparison, we vary cuts on $\tau_{21}$ and plot the $W$ efficiency versus the QCD rejection (defined as the inverse of QCD efficiency). The results, shown in Figure~\ref{fig:nsub2}, are consistent between trimmed and re-clustered jets indicating that powerful discrimination between signal and background jets using jet substructure information is possible with re-clustered jets.


Some studies have shown \cite{wbosonCMS} that un-groomed jets perform better than trimmed or otherwise groomed jets as inputs for the purposes of calculating substructure variables. Re-clustered jets have the disadvantage of always having some effective level of grooming applied (even without the additional re-clustered trimming layer), so in principle calculating substructure variables without grooming is not possible with re-clustered jets. The event display in Figure~1 highlights this issue: substructure variables are sometimes best calculated with large-area jets like those on the left hand plot, but re-clustered jets only ever appear with the reduced-area of the plot on the right. In practice, it is possible to simply associate all constituents closer than the jet parameter $R$ for the purposes of calculating shape variables. The details of such an association are left to studies within the experimental collaborations.

\subsection{Bottom-up Substructure}
\label{sec:analysis:bottomup}

An alternative to the assignment of radius $r$ constituents to the radius $R$ jets described in Sec.~\ref{sec:analysis:nsub} is to use the radius $r$ jets directly.  For example, consider the $k_t$ splitting scale\footnote{Computed by re-clustering a jet's constituents using the $k_t$ algorithm and then considering the distance metric of the last $n$ un-clusterings.} $\sqrt{d_{n,n+1}}$, which is sensitive to hard $(n+1)$-prong structure in a jet.  One can use directly the radius $r$ jets inside a radius $R$ re-clustered jet to compute $\sqrt{d_{n,n+1}}$.  If there are only two radius $r$ jets, then $\sqrt{d_{12}}$ is simply the $k_t$ distance between the radius $r$ jets.  The advantage of this approach is that there is a natural prescription for calibrations and systematic uncertainties.  The jet energy scale calibration and its uncertainties directly translate into the calibration of the bottom-up substructure variables.  Furthermore, in this approach one knows how the substructure variable calibrations and uncertainties are correlated with the re-clustered jet calibrations and uncertainties.   This information is available for the first time with this bottom-up procedure.

\begin{figure}[htbp!]
\begin{center}
\includegraphics[width=0.75\textwidth]{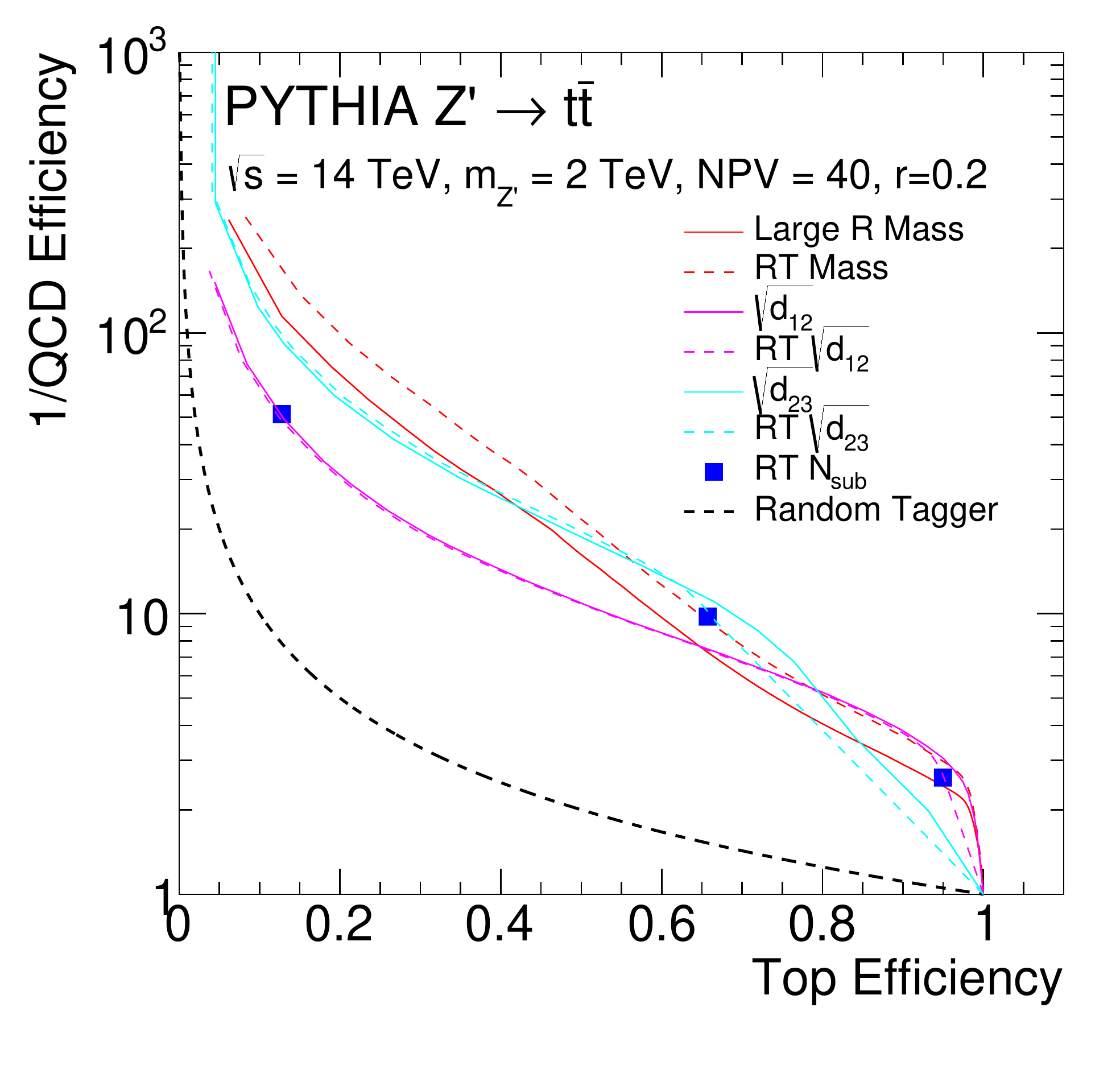}
\end{center}
\caption{The performance of a bottom-up approach to jet substructure where the radius $r$ jets are inputs to substructure variables.  Solid lines show the performance curves for large radius trimmed jets ($R_\text{sub}=0.3, f_\text{cut}=0.05$) and the dashed lines show the analogous re-clustered variable. Random tagger denotes a classifier which picks signal and background with equal probability.}
\label{fig:bottomup}
\end{figure}

Figure~\ref{fig:bottomup} compares bottom-up and top-down jet substructure variables in classifying $Z'\rightarrow t\bar{t}$ and QCD multijet events.  For the chosen parameters, the two techniques have comparable performance.  The main drawback of bottom-up substructure is that the relative efficacy depends on $p_T$ (and $r$).  When $r\gtrsim m/p_T$, or equivalently, when there are not many radius $r$ jets inside the radius $R$ jet, the experimental gains from bottom-up substructure are diminished.  For instance, if there is only one radius $r$ jet, then $\sqrt{d_{12}}=0$.  Thus, in certain kinematic regimes, bottom-up substructure may provide a powerful alternative to standard methods, but in other regimes a more dedicated analysis is required to understand correlations in calibrations and uncertainties (when jet substructure observables are built from the jet constituents).

\clearpage
\newpage

\section{Conclusion}\label{sec:conc}

Jet re-clustering is a very simple approach for creating large radius jets which emphasizes experimental flexibility by simplifying the commissioning and calibration of jet algorithms. It allows experimentalists to focus on calibrating only one jet algorithm (the small jets used as inputs to larger jets), and has a modular structure that is amenable to optimization per physics process and per event kinematics.  Uncertainties and resolutions propagate from the low level to the high level jets, simplifying the determination of mass scales and mass resolutions for large radius jets.  Pileup corrections applied to small jets, using both area-based techniques and track/vertex-based approaches, stabilize the large radius jets against pileup.  With the addition of re-clustered trimming, performance can be essentially identical to large-$R$ trimmed jets, not only for jet mass but also for shape variables such as n-subjettiness. 

Jet clustering-- both the choice of algorithm and $R$-parameter-- is an important aspect of analysis optimization in hadronic final states at the LHC.  Experimentally, it has been difficult to fully utilize such optimization because of the time and effort involved in calibrating, validating, and otherwise commissioning each separate collection of jets. Jet re-clustering provides one solution for a modular system that allows every analysis to choose the jet clustering parameters that optimize the sensitivity and accuracy of searches and measurements.

\section{Acknolwdgements}

We would like to thank Gavin Salam, Sal Rappoccio, and Hernan Reisin for useful comments on the text and Sal in particular for suggesting the title of the paper.  Additionally, we would like to thank Jon Butterworth for feedback on the manuscript.  This work is supported by the US Department of Energy (DOE) Early Career Research Program and grant DE-AC02-76SF00515. BN and MS are supported by the NSF Graduate Research Fellowship under Grant No. DGE-4747, and BN is also supported by the Stanford Graduate Fellowship.

\end{document}